\documentclass{amsart}

\usepackage{amssymb,latexsym,amsfonts,amsmath}
\usepackage{graphicx}
\usepackage{mathrsfs}
\usepackage{eso-pic}
\usepackage{epsfig}
\usepackage{graphicx}
\usepackage{color}
\usepackage{algorithm}
\usepackage{algorithmic}
\usepackage{subfigure}

\usepackage{amssymb,latexsym,amsfonts,amsmath}
\usepackage{graphicx}
\usepackage{amsthm}
\usepackage{amsmath}
\newcommand{\widebar}[1]{\overline{#1}}

\newtheorem{assumption}{Assumption}

\newtheorem{theorem}{Theorem}[section]
\newtheorem{lemma}[theorem]{Lemma}

\newtheorem{problem}[theorem]{Problem}

\theoremstyle{definition}
\newtheorem{definition}[theorem]{Definition}

\theoremstyle{remark}
\newtheorem{remark}[theorem]{Remark}
\numberwithin{equation}{section}

\newcommand{\argmin}{\textrm{arg}\min}

\begin{document}

\begin{abstract}
Finite-state abstractions (a.k.a. \emph{symbolic models}) present a promising avenue for the formal verification and synthesis of controllers in continuous-space control systems. These abstractions provide simplified models that capture the fundamental behaviors of the original systems. However, the creation of such abstractions typically relies on the availability of precise knowledge concerning system dynamics, which might not be available in many real-world applications. In this work, we introduce a novel data-driven and compositional approach for constructing finite abstractions for interconnected systems comprised of discrete-time control subsystems with partially unknown dynamics. These subsystems interact through a partially unknown static interconnection map. Our methodology for abstracting the interconnected system involves constructing abstractions for individual subsystems and incorporating an abstraction of the interconnection map. In our data-driven framework, we assume access to black-box representations of the subsystems' transition functions and a finite set of input-output pairs of the interconnection map. This allows us to collect data samples from the partially unknown subsystems, facilitating the construction of finite abstractions while ensuring their correctness. Nevertheless, the computational complexity of building finite abstractions for interconnected systems based on those of subsystems can still become formidable, especially depending on the structure of the interconnection map. To address this challenge, we introduce intermediate variables that streamline the process, breaking down the interconnection and abstraction tasks into more manageable computations. To demonstrate the effectiveness of our approach, we present results for three numerical benchmarks: the construction of a finite abstraction for a $32$-dimensional system by combining abstractions of $32$ scalar subsystems; the synthesis of a controller for an $8$-dimensional system based on its constructed finite abstraction, aimed at achieving a consensus objective; and stabilizing a network of $10$ tanks within target regions.
\end{abstract}

\title[Data-driven Construction of Finite Abstractions for Interconnected Systems: \\
A Compositional Approach]{Data-driven Construction of Finite Abstractions for Interconnected Systems: \\
A Compositional Approach}
\thanks{This work was supported by the NSF under grant CNS-2145184.}

\author[daniel ajeleye]{Daniel Ajeleye} 
\author[majid zamani]{Majid Zamani} 
\address{Department of Computer Science, University of Colorado Boulder, USA}
\email{\{daniel.ajeleye, majid.zamani\}@colorado.edu}
\urladdr{https://www.hyconsys.com/members/dajeleye/}
\urladdr{https://www.hyconsys.com/members/mzamani/}

\maketitle

\section{Introduction}\label{sec1}
Designing controllers for continuous-space systems with complex control objectives presents a formidable challenge. Fortunately, recent advances have introduced several techniques to address these obstacles. A particularly effective approach involves the creation of finite abstractions for the original control systems \cite{tabuada2009verification}. These abstractions provide concise representations of dynamical systems, designed in such a way that a discrete controller, initially constructed to enforce specific properties on the finite model, can be further refined into a hybrid controller that upholds these properties on the actual concrete system. However, the process of constructing symbolic models for large-scale systems composed of many subsystems is inherently intricate. To overcome this challenge, an effective strategy is to begin by constructing symbolic models for individual subsystems. Subsequently, through a compositional framework, one can assemble a finite abstraction of the entire network by integrating these individual abstractions. 

In this work, we introduce a novel data-driven technique for constructing a finite abstraction of a system composed of smaller interacting components, \emph{i.e.,} subsystems. In particular, our approach entails the collection of data from multiple initializations of these partially unknown subsystems, thereby enabling us to construct a data-driven finite abstraction for each individual subsystem. This abstraction process is performed concurrently and independently, as each subsystem's internal state remains isolated from the others through input-state interfaces. The interplay between these subsystems is governed by an interconnection map, which constraints the values of internal variables spanning the various subsystems. 

Furthermore, the interconnection maps are also approximated by finite abstractions. We present a compositional result, 
which demonstrates the existence of a feedback refinement relation \cite{reissig2016feedback} that bridges the abstract interconnections with the concrete ones. Drawing from insights presented in \cite{gruber2017sparsity}, which develops finite abstractions monolithically for systems with sparsely interconnected components, we have introduced a sparsity constraint to the data-driven approximation of the interconnection map. This results in a sparse matrix form of the nonlinear interconnection map. This approach facilitates the incorporation of latent variables, capturing intermediate computations, and thus breaking down the interconnection and abstraction tasks into more manageable components. Ultimately, this approach significantly reduces the overall complexity of the abstraction process. Finally, we illustrate the efficacy of our approach by applying it to three numerical benchmarks: the construction of a finite abstraction for a $32$-dimensional system by combining abstractions of $32$ scalar subsystems; the synthesis of a controller for an $8$-dimensional system based on its constructed finite abstraction, aimed at achieving a consensus objective; and stabilizing a network of $10$ tanks within specific regions. In the initial example, we demonstrate the efficiency and computational complexity of our compositional method for constructing finite abstractions of interconnected systems. Our strategy involves generating finite abstractions for the subsystems first and then integrating these abstractions to create a monolithic abstraction, as opposed to creating an abstraction for the interconnected system in its entirety. We show that the time needed to construct the abstraction with our method scales efficiently for interconnected systems with scalar-valued subsystems that exhibit nonlinear interconnections, handling dimensions up to $32$. Furthermore, the second example emphasizes synthesizing a controller to achieve consensus among subsystems linked by a linear map. The final case study spotlights the design of a controller intended to stabilize a network of $10$ tanks within a targeted region, where these models are interconnected via a nonlinear high-dimensional mapping. 

In summary, the primary contribution of this work is the construction of finite abstractions for interconnected systems using a compositional, data-driven approach with formal guarantees. By leveraging the structure of the interconnection map, we decompose the abstraction process into computationally manageable steps involving the finite abstraction of subsystems. We analyze the complexity of our approach and demonstrate, through case studies, that constructing a finite abstraction of an interconnected system in a monolithic, data-driven manner results in an abstraction size that grows exponentially with the number of subsystems. In contrast, our data-driven, divide-and-conquer strategy mitigates this challenge by significantly reducing computational complexity. Furthermore, our approach constructs finite abstractions with reduced conservatism, leading to a more effective controller domain size during synthesis and enhancing the practical utility of these abstractions.

\textbf{Related Work.} 
Limited research has been dedicated to constructing finite abstractions using data-driven methods. These findings encompass various systems, including unknown monotone systems \cite{makdesi2021efficient}, incrementally input-to-state stable control systems \cite{lavaei2022data}, and continuous-time perturbed systems \cite{kazemi2022data}. Furthermore, approaches like \cite{xue2020pac, fan2017dryvr, coppola2022data} focus predominantly on data-driven techniques for verifying unknown systems while offering probabilistic guarantees. However, it is crucial to recognize that all of these aforementioned results adopt a monolithic perspective, abstracting the entire system. It is important to acknowledge that the monolithic abstraction approach encounters scalability issues. These issues stem from the exponential increase in complexity based on the number of state variables in the model, primarily because of the gridding of state sets. As a consequence, these techniques tend to face computational limitations, especially in state spaces that exceed four dimensions.

Our approach is applicable to all classes of non-linear discrete-time control systems, in contrast to the work by \cite{makdesi2021efficient}, which is exclusively suited for monotone systems. While the result in \cite{lavaei2022data} imposes incremental stability requirements on the underlying systems, and similar to \cite{xue2020pac}, \cite{fan2017dryvr}, \cite{lavaei2023symbolic} and \cite{coppola2022data} provide finite abstractions with probabilistic guarantees, our approach stands out by offering a compositional construction of sound abstractions with a $100\%$ 
correctness guarantee. Importantly, our method does not necessitate any stability assumptions on the subsystems, nor does it impose any implicit conditions regarding the number of subsystems.

In recent years, significant progress has been made in developing a compositional framework for the construction of finite abstractions for networks of control systems. Noteworthy contributions in the literature include \cite{hussien2017abstracting, kim2018constructing, meyer2017compositional}, which construct sound symbolic models of interconnected systems in a compositional manner. Regrettably, all of these works on the compositional construction of finite abstractions require the availability of system models, which are often absent in many real-world applications. To address this challenge, one might explore various \emph{indirect data-driven} strategies to come up with models for unknown dynamical systems through identification techniques (\emph{e.g.,} \cite{ljung1998system, fattahi2018data, hou2013model} and references therein). However, obtaining an accurate model can be arduous, time-intensive, and computationally expensive in itself.

\section{Preliminaries and Definitions} \label{sec2}
\subsection{Notation}
\label{sec2_1} 
Symbols  $\mathbb{R}$, $\mathbb{R}_{>0}$, and $\mathbb{R}_{\ge0}$, respectively, represent sets of real, positive, and non-negative real numbers. Notations $\cup$, $\cap$, and $\setminus$ indicate, respectively, set union, intersection, and set difference. Similarly, $\land$ 
denotes the logical conjunction. 
The symbol $\mathbb{N}$ denotes the set of natural numbers and $\forall n\in\mathbb{N}\cup\{0\}$, symbol $\mathbb{N}_{\ge n}=\{l\in\mathbb{N}\cup\{0\}~|~l\ge n\}$. In the case where 
$a,b\in\mathbb{N}_{\ge 0}$ and $a<b$, we employ the notations $[a;b]$, $(a;b)$, $[a;b)$, and
$(a;b]$ to represent respectively the closed, open, half-open from the right, and half-open from the left intervals in $\mathbb{N}_{\ge 0}$. Alternatively, for $a,b\in\mathbb{R}$ and $a<b$, we use $[a,b]$, $(a,b)$, $[a,b)$, and
$(a,b]$ to denote the corresponding intervals in $\mathbb{R}$. For any non-empty set $Q$ and $n\in\mathbb{N}$, we denote the cardinality of $Q$ as $\mathcal{C}_d(Q)$, while $Q^n$ indicates the Cartesian product of $n$ duplicates of $Q$. Given $N$ vectors $x_i \in \mathbb R^{n_i}$, $n_i\in \mathbb N$, and $i\in\{1,\ldots,N\}$, we use $x = [x_1;\ldots;x_N]$ to denote the corresponding column vector of dimension $\sum_i n_i$. The vector $\mathbf{1}_n\in\mathbb{R}^n$ is defined as $[1;1;\dots;1]\in\mathbb{R}^{n}$, while $\mathbf{1}_{n\times n}\in\mathbb{R}^{n\times n}$ denotes an $n\times n$ matrix with all entries equal to $1$. For any $n,m,m'\in\mathbb{N}_{\ge2}$, the matrices $\mathbf{0}_{n\times m}$ and $\mathbf{I}_{n}$ represent a null matrix of dimension $n\times m$ and an identity matrix of dimension $n\times n$, respectively. Given matrices $A\in\mathbb{R}^{n\times m}$ and $B\in\mathbb{R}^{n\times m'}$, we define the augmented matrix $C:=[A;B]\in\mathbb{R}^{n\times(m+m')}$
, while $\Vert A\Vert_1$ represents the $\ell_1$-norm of $A$, \emph{i.e.,} its maximum column sum. Notation $\mathrm{dim}(Z)\in\mathbb{N}_{\ge0}$ denotes the dimension of a given set $Z$ within a vector space.  
For any $\bar p,\bar q\in \mathbb{R}^n$ and relational operator $\simeq\;\in\{\le,<,=,>,\ge\}$, where $\bar p=[p_1;\dots;p_n]$ and $\bar q=[q_1;\dots;q_n]$, $\bar p\simeq\bar  q$ is interpreted as $p_l\simeq q_l$, $\forall l\in\{1,\dots,n\}$, \emph{i.e.,} component-wise comparison. Assuming $\bar p<\bar q$, then the \emph{compact hyper-interval} $[\bar p,\bar q]$ is given as $[p_1,q_1]\times\cdots\times[p_n,q_n]$. Furthermore, given $c=[c_1;\dots;c_n]\in\mathbb{R}^{n}$, we define the sum $\oplus$ as $c\oplus [\bar p,\bar q]:=[p_1+c_1,c_1+q_1]\times\cdots\times[p_n+c_n,c_n+q_n]$. Notation $|c|$ means the entry-wise absolute value of $c\in\mathbb{R}^{n}$ \emph{i.e.,} $[|c_1|;\dots;|c_n|]$, while $\Vert c\Vert$ means the infinity norm of $c$. Similarly, $\Vert c\Vert_p$ gives the $\ell_p$-norm of $c$ for some $p\ge1$. For any $\bar r\in\mathbb{R}^n_{>0}$ and $c_0\in\mathbb{R}^n$, notation $\Phi_{\bar r}(c_0)$ is interpreted as $c_0\oplus[-\bar r,\bar r]$. For a given compact hyper-interval $H$ and discretization parameter vector $\eta_h\in\mathbb{R}^n_{> 0}$, we create a partition of $H$ into cells $\Phi_{\eta_h}(h)$ such that $H\subseteq \bigcup_{h\in[H]_{\eta_h}}\Phi_{\eta_h}(h)$, where $[H]_{\eta_h}$ represents a finite set of representative points selected from those partition sets. For any $\vartheta\in \mathbb{R}^{n\times m}$, $\Vert\vartheta\Vert$ denotes the infinity norm of $\vartheta$. 

\subsection{Discrete-Time Control Systems}
\label{sec2_2}
Here, we investigate discrete-time control systems (cf. next definition) encompassing both internal and external inputs. Internal variables facilitate interconnection with other systems, whereas external ones serve as interfaces for controllers.
\begin{definition}
\label{dt-s}
    A discrete-time control system (dt-CS) $\Xi$ is represented via a tuple 
    \begin{equation}
    \label{eq2.1}
        \Xi=(\mathcal X,U,\mathcal{U},f),
    \end{equation}
    where $\mathcal X,U$, and $\mathcal{U}$ 
    denote the state set, external input set, and internal input set 
    of the control system, respectively. These sets are assumed to be nonempty subsets of normed vector spaces with finite dimensions. The transition function, denoted as $f:\mathcal X\times U \times \mathcal{U} \rightrightarrows \mathcal X$, is a set-valued map. 
    The dt-CS $\Xi$ is characterized by difference inclusions of the following form:
    \begin{equation}
   \label{eq2.2}
       x(k+1)\in f(x(k),u(k),w(k)),
   \end{equation}
   where at time $k\in\mathbb{N}$, $x(k)\in \mathcal X$, 
   $u(k)\in U$, 
   and $w(k)\in\mathcal{U}$ represent the state, 
   external, 
   and internal input, respectively.
\end{definition}

A dt-CS $\Xi=(\mathcal X,U,\mathcal{U},f)$ 
is referred to as \emph{deterministic} when $\mathcal{C}_d(f(x, u, w))$ is at most $1$ $\forall x \in \mathcal X$, $\forall u\in U$, and $\forall w\in\mathcal{U}$. Otherwise, the dt-CS $\Xi$ is considered \emph{non-deterministic}. In addition, $\Xi$ is classified as \emph{finite} if $\mathcal X$, $U$, and $\mathcal{U}$ are finite sets, and $\Xi$ is said to be \emph{infinite} otherwise.

\begin{remark}
\label{rmk2.2}
If a dt-CS $\Xi$ does not have internal inputs
, the tuple \eqref{eq2.1} in Definition \ref{dt-s} simplifies to a \emph{simple system}
\begin{equation}
    \label{eq2.3}
        \Xi=(X,U,f),
    \end{equation}
where the set-valued map $f$ becomes $f: X \times U \rightrightarrows X$. Consequently, the formulation in \eqref{eq2.2} is reduced to:
\begin{equation}
   \label{eq2.4}
       x(k+1)\in f(x(k),u(k)).
   \end{equation}
\end{remark}

Subsequently, we utilize the notion of the dt-CS in equations \eqref{eq2.3} and \eqref{eq2.4}, to specifically refer to an interconnected dt-CS. This dt-CS is constructed by interconnecting several subsystems described as in \eqref{eq2.1} and \eqref{eq2.2}. Next, we introduce the notion of feedback refinement 
relation, which provides a relation between two dt-CSs in terms of controller synthesis. 

\subsection{Feedback Refinement Relations}
\label{sec2_3}
Here, we recall the notion of feedback refinement relations \cite{reissig2016feedback} that establish a relationship between two dt-CSs, as described in Definition \ref{dt-s} and quantify the relationship between them in terms of controller synthesis \cite{tabuada2009verification}.
\begin{definition}
    \label{asf}
    Consider two dt-CS $\Xi_i=(\mathcal X_i,U_i,\mathcal{U}_i,f_i)$
    , where $i\in\{1,2\}$, such that $U_2\subseteq U_1$. 
    There is a feedback refinement relation from $\Xi_1$ to $\Xi_2$ if there exist nonempty relations $\mathcal Q\subseteq\mathcal X_1\times \mathcal X_2$ and $\mathcal R\subseteq\mathcal U_1\times\mathcal U_2$ such that, $\forall(x_1,x_2)\in \mathcal Q$ $\forall w_1\in \mathcal U_1$ $\exists w_2\in\mathcal U_2$ such that $(w_1,w_2)\in\mathcal R$ and the next two conditions hold:
    \begin{itemize}
        \item $U_{\Xi_2}(x_2,w_2)\subseteq U_{\Xi_1}(x_1,w_1)$, where $U_{\Xi_i}(x_i,w_i):=\{u_i\in U_i~|~f_i(x_i,u_i,w_i)\neq\emptyset\}$ is the set of \emph{admissible} external inputs for state $x_i\in\mathcal X_i$ and internal input $w_i\in\mathcal U_i$, $\forall i\in\{1,2\}$; 
        \item if $u\in U_{\Xi_2}(x_2,w_2)$, then $f_1(x_1,u,w_1)\times f_2(x_2,u,w_2)\subseteq \mathcal Q$.
    \end{itemize}
\end{definition}

Furthermore, whenever there is a feedback refinement relation 
as in Definition \ref{asf} from dt-CS $\Xi_1$ to $\Xi_2$, we denote it by $\Xi_1\preceq_\mathcal{Q}\Xi_2$. 
When there are no internal inputs
, Definition \ref{asf} simplifies to the next one.
\begin{definition}
    \label{asf_mono}
    Consider two dt-CSs $\Xi_i=(X_i,U_i,f_i)$
    , where $i\in\{1,2\}$, such that $U_2\subseteq U_1$. 
    There is a feedback refinement relation from $\Xi_1$ to $\Xi_2$ if there exists a nonempty relation $ Q\subseteq X_1\times X_2$ 
    such that, $\forall(x_1,x_2)\in Q$ 
    the following two conditions hold:
    \begin{itemize}
        \item $U_{\Xi_2}(x_2)\subseteq U_{\Xi_1}(x_1)$, where $U_{\Xi_i}(x_i):=\{u_i\in U_i~|~f_i(x_i,u_i)\neq\emptyset\}$ is the \emph{admissible} input set for state $x_i\in X_i$
        , $\forall i\in\{1,2\}$; 
        \item if $u\in U_{\Xi_2}(x_2)$, then $f_1(x_1,u)\times f_2(x_2,u)\subseteq Q$.
    \end{itemize}
    
\end{definition}

\begin{remark}
   The feedback refinement relations introduced in \cite{reissig2016feedback} address two key limitations of the abstraction and refinement process based on alternating simulation relations and related concepts \cite{tabuada2009verification}. Specifically, they eliminate the need for full state information by allowing the use of quantized information and mitigate the complexity of refinement by avoiding the requirement to use finite abstractions as a building block within the controller. For further details on these aspects, we refer the interested reader to \cite{reissig2016feedback}. Accordingly, we have adopted the notion of feedback refinement relations as the system relation in this work.
\end{remark}

\section{Systems Interconnection}
\label{sec3}
In this section, our focus lies in the analysis of interconnected systems, comprised of several dt-CSs.
 
\subsection{Interconnected dt-CSs}
\label{sec3.1}
Now, we define an interconnection of several dt-CSs.

\begin{definition}
    \label{intercon}
    Consider $N\in\mathbb{N}_{\ge 1}$ dt-CSs $\Xi_i=(\mathcal X_i,U_i,\mathcal{U}_i,f_i)$
    , where $i\in[1;N]$. An interconnection $\mathcal{I}$ between the $N$ subsystems is defined by the tuple $\mathcal{I}=(\prod_{i=1}^{N} \mathcal{X}_{i},\prod_{i=1}^N \mathcal{U}_i,\mathcal{M})$, where $\mathcal{M}$ is the interconnection map and defined as, 
    \begin{equation}
        \label{inter_map}
        \mathcal{M}:\prod_{i=1}^{N} \mathcal{X}_{i}\rightarrow\prod_{i=1}^N \mathcal{U}_i.
    \end{equation}
\end{definition}

For the sake of simple presentation, an interconnection $\mathcal{I}$ as in Definition \ref{intercon} is concisely written as tuple $\mathcal{I}=(\mathcal{X}^\mathcal{I},\mathcal{U}^\mathcal{I},\mathcal{M})$ where $\mathcal{X}^\mathcal{I}=\prod_{i=1}^{N} \mathcal{X}_{i}$ and $\mathcal{U}^\mathcal{I}=\prod_{i=1}^N \mathcal{U}_i$.
Next, we define the interconnected dt-CS.
\begin{definition}
    \label{intercon2}
    Consider a collection of $N\in\mathbb{N}_{\ge 1}$ subsystems $\Xi_i=(\mathcal X_i,U_i,\mathcal{U}_i,f_i)$
    , where $i\in [1; N]$, along with an interconnection $\mathcal{I}=(\mathcal{X}^\mathcal{I},\mathcal{U}^\mathcal{I},\mathcal{M})$ that defines the coupling among these subsystems. The interconnected dt-CS $\Xi = (X, U, f)$ 
    is denoted by $\mathcal{I}(\Xi_1,\ldots, \Xi_N)$, and defined as follows: 
    \begin{itemize}
        \item $X=\prod_{i=1}^N \mathcal X_i$ and $U=\prod_{i=1}^N U_i$
        ;
        \item for any state $x=[x_1;\ldots;x_N]\in X$ and input \\$u=[u_1;\ldots;u_N]\in U$, it holds that
    \end{itemize}
    \begin{equation}
    \label{sub_f}
            f(x,u):= \{[x_1';\ldots;x_N']~|~x_i'\in f_i(x_i,u_i,w_i)~\forall i\in[1;N]\},
        \end{equation}
        where $[w_1;\ldots;w_N]=\mathcal{M}([x_1;\ldots;x_N])$.
\end{definition}

    Definitions \ref{intercon} and \ref{intercon2} highlight the role of internal inputs 
    within a network of dt-CSs. These internal variables facilitate the construction of independent abstractions for subsystems. Moreover, the interconnection map allows for the decomposition of complex interconnected systems into simpler, more manageable components, enabling modular analysis and control. This approach allows each subsystem to be addressed individually while preserving coordinated behavior, thereby improving scalability.

The following definition introduces a notion of abstract interconnection.
\begin{definition}
    \label{inter_abs}
    Consider interconnections $\mathcal{I}_i=(\mathcal{X}^{\mathcal{I}_i},\mathcal{U}^{\mathcal{I}_i},\mathcal{M}_i)$, where $i\in\{1,2\}$ such that ${\mathrm{dim}(\mathcal{X}^{\mathcal{I}_1})}={\mathrm{dim}(\mathcal{X}^{\mathcal{I}_2})}$ and ${\mathrm{dim}(\mathcal{U}^{\mathcal{I}_1})}={\mathrm{dim}(\mathcal{U}^{\mathcal{I}_2})}$. Interconnection $\mathcal{I}_2$ is called an abstraction 
    of $\mathcal{I}_1$ if there exist relations $\tilde Q\subseteq \mathcal{X}^{\mathcal{I}_1}\times \mathcal{X}^{\mathcal{I}_2}$ and $\tilde{ \mathcal{R}}\subseteq \mathcal{U}^{\mathcal{I}_1}\times\mathcal{U}^{\mathcal{I}_2}$ such that $\forall(x,\hat x)\in\tilde Q$ $\forall w\in\mathcal{M}(x)$ $\exists\hat w\in\hat{\mathcal{M}}(\hat x)$ such that $(w,\hat w)\in\tilde{\mathcal{R}}$.
\end{definition}

The mentioned interconnection abstraction is  employed to construct abstractions of interconnected dt-CSs using the subsystems' abstractions compositionally. Next subsection introduces the main problem which we aim at solving.

\subsection{Problem Formulation}
\label{sec4.0}
Here, we examine \( N \) continuous-space, deterministic discrete-time control systems (dt-CSs), denoted by \( \Xi_i \), as described in \eqref{eq2.1} and \eqref{eq2.2}, where \( i \in [1; N] \). The transition maps $f_i$ in \eqref{eq2.2} are considered to be \emph{partially unknown} (cf. Assumption \ref{assume2}). Furthermore, these $N$ subsystems are interconnected via an interconnection map $\mathcal{M}$ as in Definition \ref{intercon} which is considered to be \emph{partially unknown} as well (cf. Assumption \ref{assume3}).

The primary objective of this work is to construct a finite abstraction of an interconnection of $N$ dt-CSs $\Xi_i$, for the sake of synthesizing controllers  using the constructed abstraction. In our setting, although the underlying dynamics of \( \Xi_i \) (denoted by \( f_i \) in \eqref{eq2.2}) are partially unknown, we have access to a black-box representation of the systems, allowing us to collect data points from their trajectories. We gather these data samples in a set $\mathcal{D}_{\mathcal{N}_{C_i}}\!:=\!\{(x_l,u_l,w_l,x_l')~|~x_l'=f_i(x_l,u_l,w_l)\text{ where }x_l\in \mathcal{X}_i,~u_l\in U_i,\text{ and }w_l\in\mathcal{U}_i,\,l\!\in\![1;\mathcal{N}_{C_i}]\}$. Note that we abuse notation by using $x_l$ and $w_l$ later to represent the data point for the state of the interconnected dt-CSs and the vector containing all internal inputs, respectively (cf. Subsection \ref{sec4.2}). Although the interconnection map in \eqref{inter_map} is assumed to be a partially unknown nonlinear map and the specific connections between subsystems are unknown, we do have access to a finite set of input-output pairs for the interconnection map. 
 
These collected data-points are stored in the set $\mathcal{D}_{\mathcal{N}_I}:=\{(x_l,w_l)\in\mathcal{X}^\mathcal{I}\times\mathcal{U}^\mathcal{I}~|~w_l=\mathcal{M}(x_l),\,l\in[1;\mathcal{N}_I]\}$. Under these assumptions, we formulate the main problem that we aim to address in this work.
\begin{problem}
    \label{prob1}
    Consider an interconnected dt-CS $\Xi$ as in Definition \ref{intercon2} composed of $N$ subsystems $\Xi_i$, $i\in[1;N]$, where transition maps $f_i$ and the interconnection map $\mathcal{M}$ are partially unknown. Develop a compositional, data-driven approach based on the sets of data $\mathcal{D}_{\mathcal{N}_{C_i}}$, $i\in[1;N]$, and $\mathcal{D}_{\mathcal{N}_I}$, to construct a finite abstraction $\widehat\Xi$ such that $\Xi\preceq_Q \widehat\Xi$, where $Q$ is a feedback refinement relation.
\end{problem}
In the next section, we introduce a data-driven approach for constructing abstractions of subsystems within a network and the interconnection abstraction as in Definition \ref{inter_abs}.

\begin{figure}[!ht]
    \centering
    \includegraphics[width=15.0cm]{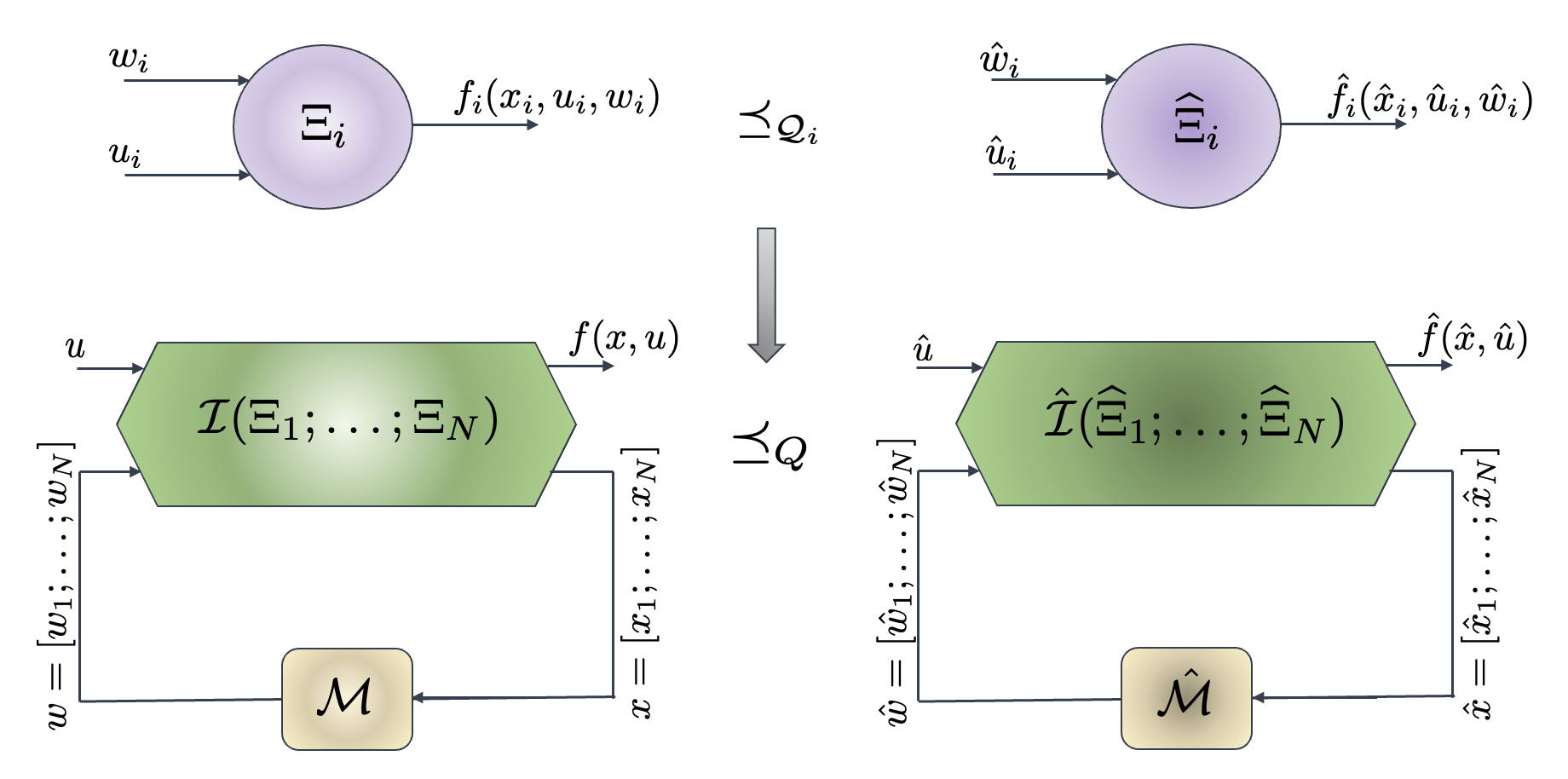}
    \caption{Interconnection of $N$ finite abstractions $\widehat\Xi_i$ while maintaining the feedback refinement relation $Q$. This interconnection is achieved by leveraging the compositionality result (cf. Theorem \ref{asf_thrm}).} 
    \label{fig1}
\end{figure}

\section{Data-Driven Construction of Finite Abstractions} 
\label{sec4}
 In this section, we tackle Problem \ref{prob1} in a step-by-step fashion, commencing with the creation of finite abstractions for individual subsystems. Then, we develop a finite approximation of the interconnection. Finally, we integrate these constructed subsystem abstractions and the interconnection approximation to construct an overall abstraction, encompassing the behavior of the entire interconnected system.  
\subsection{Symbolic Abstractions for Subsystems}
\label{sec4.1}
Consider $N$ dt-CSs $\Xi_i=(\mathcal{X}_i,U_i,\mathcal{U}_i,f_i)$
, where $i\in[1;N]$, which are interconnected by a map $\mathcal{M}$.
To tackle Problem \ref{prob1}, we first raise the following assumptions.

\begin{assumption}
    \label{assume2} For any $u\in U_i$, the transition map $f_i$ is locally Lipschitz continuous with respect to $x$ and $w$, with Lipschitz constants $\mathcal{L}_{x_i}(u)$ and $\mathcal{L}_{w_i}(u)$, respectively, $\forall i\in[1;N]$, \emph{i.e.,} for any $\hat x\in \mathcal{X}_i$ and $\hat w\in \mathcal{U}_i$ and neighborhoods $\Phi_{\eta_w/2}(
    \hat x)$ and $\Phi_{\eta_w/2}(
    \hat w)$ of $\hat x$ and $\hat w$, respectively, one has
    \begin{equation}
    \begin{split}
    \Vert f_i(x_1^i,\!u,\!w_1^i)-f_i(x_2^i,\!u,\!w_2^i)\Vert\leq & \;\mathcal{L}_{x_i}(u)\Vert x_1^i-x_2^i\Vert\\
    &+\mathcal{L}_{w_i}(u)\Vert w_1^i-w_2^i\Vert,
    \end{split}
\end{equation}
    $\forall x_1^i,x_2^i\in\Phi_{\eta_x/2}(
    \hat x)$ and $\forall w_1^i,w_2^i\in\Phi_{\eta_w/2}(
    \hat w)$.  
    In addition, we assume that the Lipschitz constants $\mathcal{L}_{x_i}(u)$ and $\mathcal{L}_{w_i}(u)$ are given.
    \end{assumption}
\begin{assumption}
\label{assume3}
    The interconnection map $\mathcal{M}$ is Lipschitz continuous with a Lipschitz constant $\mathcal{L}_\mathcal{M}$, \emph{i.e.,} for any $x_1,x_2\in \mathcal{X}^\mathcal{I}$, 
    one has
    \begin{equation}
    \begin{split}
    \Vert \mathcal{M}(x_1)-\mathcal{M}(x_2)\Vert\leq & \;\mathcal{L}_{\mathcal{M}}\Vert x_1-x_2\Vert.
    \end{split}
\end{equation} 
Additionally, we assume the Lipschitz constant $\mathcal{L}_\mathcal{M}$ is given.
\end{assumption}

\begin{remark}
\label{rmk_assump}
Note that the results presented in \cite{wood1996estimation} can be employed to estimate the Lipschitz constants mentioned in Assumption \ref{assume2} and \ref{assume3} for the unknown maps (cf. Appendix \ref{apdx1}). When the estimates of the Lipschitz constants are unavailable, they are typically inferred from finite data sets obtained from the black-box representations of the unknown subsystem transition functions and the interconnection map. However, in this work, we assume that the estimates of the Lipschitz constants are known and do not consider any confidence regarding these estimations. 
\end{remark}

For simplicity, in the remainder of this subsection we omit the index $i$ from the representations of the subsystems. Let us denote by $\eta_x\in(\mathbb{R}_{>0})^{\mathrm{dim}(\mathcal{X})}$, $\eta_u\in(\mathbb{R}_{>0})^{\mathrm{dim}(U)}$ and $\eta_w\in(\mathbb{R}_{>0})^{\mathrm{dim}(\mathcal{U})}$ the discretization parameters for $\mathcal{X}$, $U$, and $\mathcal{U}$, respectively. Hence, we generate symbolic state set $\hat{\mathcal{X}}:=[\mathcal{X}]_{\eta_x}$, $\hat U:=[U]_{\eta_u}$, and $\hat{\mathcal{U}}:=[\mathcal{U}]_{\eta_w}$. Accordingly, the exact reachable set of states from a grid cell centered at $\hat x\in\hat{\mathcal{X}}$, under an internal input grid cell centered at $\hat w\in\hat{\mathcal{U}}$, and under an external input $\hat u\in \hat U$, is defined as $\mathbf{R}(\hat x,\hat u,\hat w):=\{f(x,\hat u, w)~|~x\in\Phi_{\eta_x/2}(\hat x) \text{ and } w\in\Phi_{\eta_w/2}(\hat w)\}$. Our objective is to construct an over-approximation of $\mathbf{R}(\hat x,\hat u,\hat w)$ for any $(\hat x,\hat u,\hat w)\in\hat{\mathcal{X}}\times \hat U\times\hat{\mathcal{U}}$ using a function called \emph{growth bound}, formally defined as follows.
\begin{definition}
    \label{kappa_C}
    Given a dt-CS $\Xi=(\mathcal{X},U,\mathcal{U},f)$ 
    with corresponding symbolic sets $\hat{\mathcal{X}}$, $\hat U$, and $\hat{\mathcal{U}}$, a function $\kappa:\mathbb{R}^{\mathrm{dim}(\mathcal{X})}_{\ge0}\times\hat{\mathcal{X}}\times \hat U\times\hat{\mathcal{U}}\rightarrow\mathbb{R}^{\mathrm{dim}(\mathcal{X})}_{\ge0}$ satisfying
    \begin{equation}
        \label{kap_c_eqn}
        |x_1'-x_2'|\leq \kappa(|x_1-x_2|,\hat x,\hat u,\hat w),
    \end{equation}
where $x_1'= f(x_1,\hat u,w_1)$ and $x_2'= f(x_2,\hat u,w_2)$, for any $x_1,x_2\in\Phi_{\eta_x/2}(\hat x)$, $w_1,w_2\in\Phi_{\eta_w/2}(\hat w)$, is called a \emph{growth bound} of $\Xi$.
\end{definition}
We now formally define finite abstractions of subsystems.
\begin{definition}
    \label{sub_abs}
    Given a dt-CS $\Xi=(\mathcal X, U,\mathcal U, f)$ and a growth bound $\kappa$, a finite dt-CS $\widehat\Xi=(\hat{\mathcal X}, \hat U, \hat{\mathcal U}, \hat f)$ is a finite abstraction of $\Xi$, with the transition map $\hat f:\hat {\mathcal X}\times\hat U\times\hat{\mathcal U}\rightrightarrows\hat{\mathcal X}$ if for any $\hat x\in\hat{\mathcal X}$, $\hat w\in\hat{\mathcal U}$ and $\hat u\in \hat U$, $\hat x'\in\hat{f}(\hat x,\hat u,\hat w)$ for all $\hat x'\in\hat{\mathcal X}$ where $(f(\hat x,\hat u,\hat w)\oplus[-q',q'])\cap\Phi_{\eta_x/2}(\hat x')\neq\emptyset$, with $q'=\kappa(\eta_x,\hat x,\hat u,\hat w)$.
\end{definition}
According to Definition \ref{sub_abs}, in the abstraction, a transition from a discrete state under a given discrete input corresponds to the set of discrete states whose grid cells overlap with the over-approximation of the reachable states of the concrete systems, as determined by the growth bound function. In addition, we present the following theorem, which shows the usefulness of finite abstractions in Definition \ref{sub_abs} by establishing a feedback refinement relation (namely set membership relation) between a concrete dt-CS and its finite abstraction. Establishing a feedback refinement relation \cite{reissig2016feedback} between a dt-CS and its finite
abstraction ensures that any robust controller designed for the abstraction, which enforces a specification (\emph{e.g.},
LTL \cite{baier2008principles}, $\omega$-regular properties described by universal co-B\"uchi automata \cite{ajeleye2024data}, reach-avoid \cite{ajeleye2022output},
\emph{e.t.c.}) can be refined back to the dt-CS to achieve the same specification.
\begin{theorem}
    \label{sub_abs_thrm}
    Consider a dt-CS $\Xi$ as in Definition \ref{dt-s}, and let $\widehat\Xi=(\hat{\mathcal X}, \hat U, \hat{\mathcal U}, \hat f)$ be its finite abstraction according to Definition \ref{sub_abs}. Then $\Xi\preceq_{\mathcal{Q}}\widehat\Xi$, where the feedback refinement relation $\mathcal{Q}$ is defined as $(x,\hat x)\in\mathcal{Q}$ if $x\in\Phi_{\eta_x/2}(\hat x)$. 
\end{theorem}

The proof follows a structure similar to that of \cite[Theorem VIII.4]{reissig2016feedback}. For the sake of completeness, the full proof is provided in Appendix \ref{apdx3}.

  Remark that a version of the growth bound satisfying \eqref{kap_c_eqn} is introduced in \cite{reissig2016feedback} for model-based and continuous-time settings. However, considering scenarios where the subsystem dynamics is unknown and to reduce the dependency on the subsystem dynamics, we adopt a parametrized function \cite{kazemi2022data,ajeleye2023data} presented in \eqref{k_c_para}, as a candidate growth bound. 

For any $\bar r\in\mathbb{R}^{\mathrm{dim}(\mathcal{X})}_{\ge0}$, $\hat x\in\hat{\mathcal{X}}$, $\hat u\in\hat U$ and $\hat w\in\hat{\mathcal{U}}$, we present the parametrized candidate growth bound as follows:
\begin{equation}
    \label{k_c_para}
    \kappa_\vartheta(\bar r,\hat x,\hat u,\hat w):=\vartheta_1(\hat x,\hat u,\hat w)\bar r+\vartheta_2(\hat x,\hat u,\hat w),
\end{equation}
where $\vartheta_1\in\mathbb{R}_{\ge0}^{\mathrm{dim}(\mathcal{X})\times\mathrm{dim}(\mathcal{X})}$, $\vartheta_2\in\mathbb{R}_{\ge0}^{\mathrm{dim}(\mathcal{X})}$, and $\vartheta\in\mathbb{R}_{\ge0}^{p}$ is a column vector formed by stacking those of $\vartheta_1$ and $\vartheta_2$, with $p=\mathrm{dim}(\mathcal{X})(\mathrm{dim}(\mathcal{X})+1)$.
Remark that for every abstract state, parameters of $\vartheta_1$ and $\vartheta_2$ are locally defined, resulting in a less conservative growth bound. The following lemma, with its proof provided in Appendix \ref{apdx2}, presents a candidate growth bound. 
\begin{lemma}
\label{lema1_c}
          Consider a pair \( (\hat{u}, \hat{w}) \in \hat{U} \times \hat{\mathcal{U}} \). By Assumption \ref{assume2}, suppose that for any \( x_1, x_2 \in \Phi_{\eta_x/2}(\hat{x}) \) and \( w_1, w_2 \in \Phi_{\eta_w/2}(\hat{w}) \),
\begin{equation}
    \label{eqlp2}
    \begin{split}
    \Vert f(x_1, \hat{u}, w_1) - f(x_2, \hat{u}, w_2) \Vert \leq & \; \mathcal{L}_x(\hat{u}) \Vert x_1 - x_2 \Vert \\
    & + \mathcal{L}_w(\hat{u}) \Vert w_1 - w_2 \Vert.
    \end{split}
\end{equation}
Then, by setting \( \vartheta_1(\hat{x}, \hat{u}, \hat{w}) = \mathcal{L}_x(\hat{u}) \mathbf{1}_{\mathrm{dim}(\mathcal{X}) \times \mathrm{dim}(\mathcal{X})} \) and \( \vartheta_2(\hat{x}, \hat{u}, \hat{w}) = \mathcal{L}_w(\hat{u}) \eta_w \) in \eqref{k_c_para}, for all \( \hat{x} \in \hat{\mathcal{X}}, \hat{u} \in \hat{U}, \text{ and } \hat{w} \in \hat{\mathcal{U}} \), this yields a candidate growth bound.
       \end{lemma}
\begin{remark}
    \label{rem_conserve}
    Note that our objective is to identify a candidate growth bound, \( \kappa_\vartheta \), in the form of \eqref{k_c_para}, which is less conservative than the one proposed in Lemma \ref{lema1_c}. Specifically, we seek a growth bound that satisfies:  
\begin{equation}
    \label{eq9a}
    \Vert \vartheta_1(\hat{x}, \hat{u}, \hat{w}) \Vert \leq \mathcal{L}_x(\hat{u}) \quad \text{and} \quad \Vert \vartheta_2(\hat{x}, \hat{u}, \hat{w}) \Vert \leq \mathcal{L}_w(\hat{u}) \Vert \eta_w \Vert,
\end{equation}  
(cf. the objective functions and constraints of the optimization problem in \eqref{c_rcp} and \eqref{c_scp}). However, using a candidate growth bound as proposed in Lemma \ref{lema1_c}, or any larger bound, may result in a finite abstraction that is overly conservative, as illustrated in the case study in Subsection \ref{sec53}.
\end{remark}
We now present a data-driven approach utilizing data set $\mathcal{D}_{\mathcal{N}_C}$ to compute a candidate growth bound as in \eqref{k_c_para}. Our approach also offers a formal correctness guarantee for \eqref{k_c_para} implying that it is a growth bound for dt-CS in \eqref{eq2.2} (cf. Theorem \ref{c_unknown_data}). The primary objective is to seek a growth bound that is less conservative in terms of over-approximating the reachable sets. In our proposed framework, we initially formulate the candidate growth bound in \eqref{k_c_para} as the following robust convex program:
\begin{equation}
   \label{c_rcp}
   \text{RCP:}
   \begin{cases}
       \min_{\vartheta} \;\;\;\;\mathbf{1}_{p}^\top\vartheta&\\
       \text {{\bf s.t.}}~\vartheta\in[0,\Bar{\vartheta}],\forall x_1,x_2\in\Phi_{\eta_x/2}(\hat{x}),&\\~\;\;\;\forall w_1,w_2\in\Phi_{\eta_w/2}(\hat w),&\\
       ~\;\;\;|x'_1\!-\!x'_2|\!-\!\kappa_{\vartheta}(|x_1\!-\!x_2|,\hat{x},\!\hat{u},\!\hat{w})\le 0,&
   \end{cases}
   \end{equation}
where $x_1'= f(x_1,\hat u,w_1)$, $x_2'= f(x_2,\hat u,w_2)$, $p=\mathrm{dim}(\mathcal{X})(\mathrm{dim}(\mathcal{X})+1)$, $\vartheta\in\mathbb{R}_{\ge0}^{p}$ is a column vector formed by stacking those of $\vartheta_1$ and $\vartheta_2$, and $\Bar{\vartheta}:=\min\{\mathcal{L}_x(\hat u),\,\mathcal{L}_w(\hat u)\Vert\eta_w\Vert\}\mathbf{1}_p$.

\begin{figure}[!ht]
\vspace{-0.2cm}
    \centering
    \includegraphics[width=15.0cm]{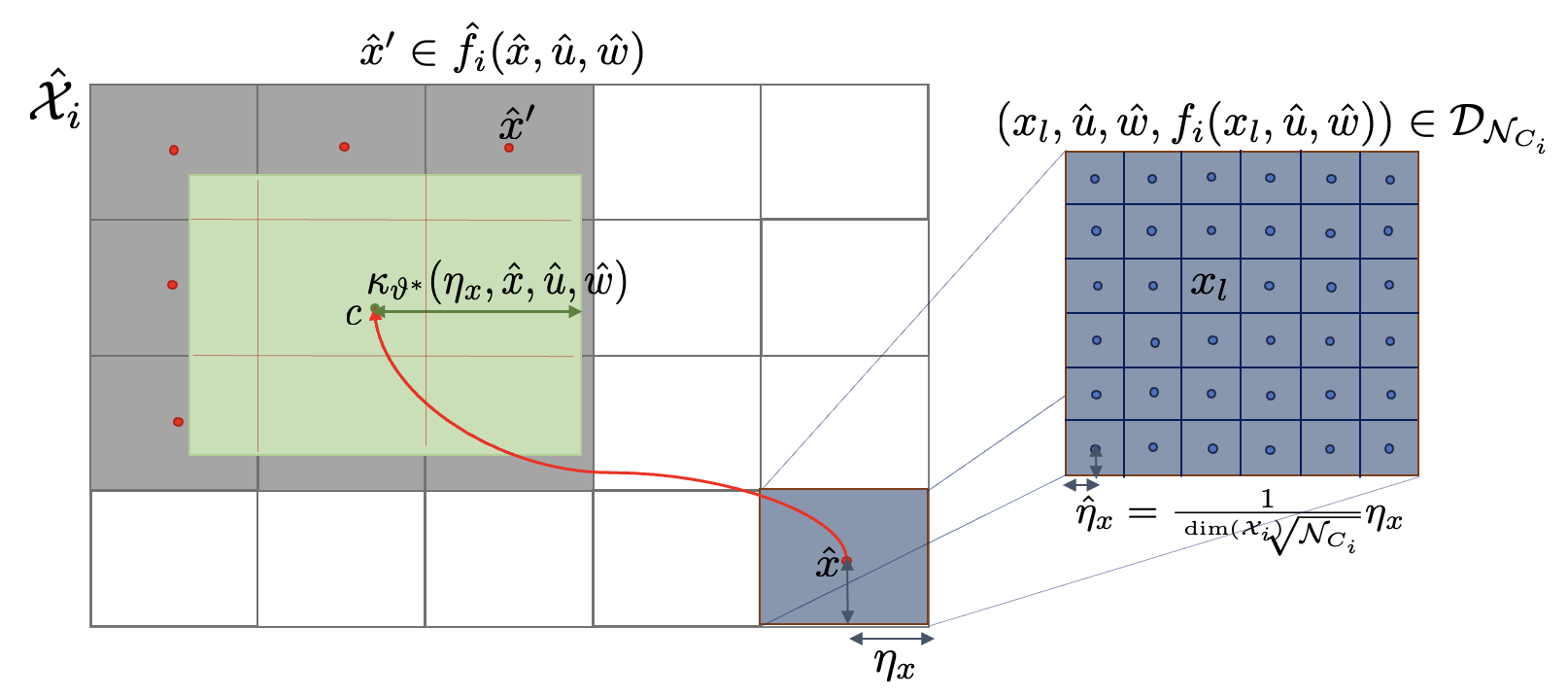}
    \caption{A 2-dimensional depiction of a finite abstraction for a subsystem, constructed using Algorithm \ref{alg_sa}.    
    }
    \label{fig2}
\end{figure}

One can readily verify that a feasible solution of the RCP in \eqref{c_rcp} provides a growth bound as in \eqref{kap_c_eqn} for dt-CS in \eqref{eq2.2}. Unfortunately, a precise knowledge of the dynamic is required for solving the problem. To resolve this issue, we collect data samples from trajectories of partially unknown dt-CS and propose a scenario convex program (SCP) corresponding to the original RCP. The scenario-based optimization approach in \cite{calafiore2006scenario} is a Probably Approximately Correct (PAC) statistical learning framework that provides probabilistic guarantees on solution quality based on a finite number of independent and identically distributed (\emph{i.i.d.}) samples drawn from an uncertainty space. In our work, we leverage the connection between robust convex programming and its scenario-based variant, primarily using grid-based data samples (deterministic sampling). The grid-based sampling approach, grounded in Lipschitz continuity assumptions, provides a \( 100\% \) guarantee of the soundness of the main analysis outcome. In contrast, PAC-based approaches offer only probabilistic guarantees. However, PAC-based methods can achieve significantly lower sample complexity compared to our grid-sampling approach. Grid-sampling methods have also been employed to solve robust programs for partially unknown systems in other contexts, as demonstrated in \cite{salamati2022data, salamati2021data, ajeleye2023data, ajeleye2024data, nejati2023formal}. To achieve this, consider a set of $\mathcal{N}_C$ data points $\mathcal{D}_{\mathcal{N}_C}$ collected within cells $\Phi_{\hat\eta_x/2}(\tilde x)$ where $\tilde x\in[\Phi_{\eta_x/2}(\hat x)]_{\hat \eta_x}$, which are sub-grids themselves within the cell $\Phi_{\eta_x/2}(\hat x)$, where $\hat\eta_x:=\dfrac{1}{\sqrt[\mathrm{dim}(\mathcal{X})]{\mathcal{N}_C}} \eta_x$ (cf. Fig. \eqref{fig2}). The proposed size of the sub-grid cells is due to extracting $\mathcal{N}_C$ data points from the primary cell $\Phi_{\eta_x/2}(\hat x)$, which has a dimension of $\mathrm{dim}(\mathcal{X})$. By leveraging the data set $\mathcal{D}_{\mathcal{N}_C}$, for any $(\hat x,\hat u,\hat w)\in\hat{\mathcal{X}}\times \hat U\times\hat{\mathcal{U}}$, we propose the SCP associated to the RCP in \eqref{c_rcp} for a cell $\Phi_{\eta_x/2}(\hat x)$ as
\begin{equation}
   \label{c_scp}
   \text{SCP:}
   \begin{cases}
       \min_{\vartheta} \;\;\;\;\mathbf{1}_{p}^\top\vartheta&\\
       \text {{\bf s.t.}}~~\vartheta\in[0,\Bar{\vartheta}],~\forall~l,\bar l\in\{1,\ldots,\mathcal{N}_C\}, 
       \\
       \;\;|x'_l\!-\!x'_{\bar l}|\!-\!\vartheta_1\!(\hat{x},\hat{u},\hat{w})|x_l\!-\!x_{\bar l}|\!-\!\vartheta_2(\hat{x},\hat{u},\hat{w})\!+\!\varrho\!\le\! 0,&
   \end{cases}
   \end{equation}
   where $\varrho\in\mathbb{R}^{\mathrm{dim}(\mathcal{X})}_{\ge0}$ can be obtained as outlined in the following theorem.

\begin{theorem}
    \label{c_unknown_data}
    Consider a dt-CS $\Xi=(\mathcal{X},U,\mathcal{U},f)$
    . For any $(\hat{x},\hat{u},\hat{w})\in [\mathcal{X}]_{\eta_x}\times [U]_{\eta_u}\times[\mathcal{U}]_{\eta_w}$, suppose for a cell $\Phi_{\eta_x/2}(\hat{x})$, $[\Phi_{\eta_x/2}(\hat{x})]_{\hat{\eta}_x}$ is constructed where $\hat\eta_x:=\dfrac{1}{\sqrt[\mathrm{dim}(\mathcal{X})]{\mathcal{N}_C}} \eta_x$. Then, the solution of \eqref{c_scp} provides a growth bound as in \eqref{kap_c_eqn} corresponding to $(\hat{x},\hat{u},\hat{w})$, where
    \begin{equation}
        \label{gamma_c}
\varrho:=4\mathcal{L}_x(\hat u)\hat\eta_x+2\mathcal{L}_w(\hat u)\Vert{\eta}_w\Vert\mathbf{1}_{\mathrm{dim}(\mathcal{X})},
    \end{equation} 
    $\mathcal{L}_x(\hat u)$ and $\mathcal{L}_w(\hat u)$ are Lipschitz constants as in Assumption \ref{assume2}.
\end{theorem}

   \begin{proof}
       \label{prf1_c} 
       It can be readily verified that the optimization problem \eqref{c_scp} admits a feasible solution. For any fixed $(\hat{x},\hat{u},\hat w)\in[\mathcal{X}]_{\eta_x}\times [U]_{\eta_u}\times[\mathcal{U}]_{\eta_w}$, let 
    \begin{equation*}
        \begin{split}
        \beta(\vartheta,x_1,x_2,w_1,w_2):=&~\!|f(x_1,\hat{u},w_1)-f(x_2,\hat{u},w_2)|\\&-\vartheta_1(\hat x,\hat u,\hat w)|x_1-x_2|-\vartheta_2(\hat x,\hat u,\hat w),\\
        \end{split}
    \end{equation*}
    for any $x_1,x_2\in\Phi_{\eta_x/2}(\hat{x})\text{ and }w_1,w_2\in\Phi_{\eta_w/2}(\hat{w})$.
    In addition, let $\vartheta^{*}$ be the optimal solution of SCP \eqref{c_scp}.
       By considering $x_1,x_2\in\Phi_{\eta_x/2}(\hat x)$ and picking samples $x_l,x_{\bar l}$ from cells $\Phi_{\hat{\eta}_x/2}(x_l),\Phi_{\hat{\eta}_x/2}(x_{\bar l})\subset\Phi_{\eta_x/2}(\hat{x})$, and $w_1',w_{2}'\in\Phi_{\eta_w/2}(\hat{w})$, one gets 
       \begin{equation*}
       \begin{aligned}
       \Vert\beta&(\vartheta^{*},x_1,x_2,w_1,w_2)-\beta(\vartheta^{*},x_l,x_{\bar l},w_1',w_{2}')\Vert\\
       \le&\,\Vert f(x_1,\hat{u},w_1)-f(x_l,\hat{u},w_1')\Vert +\Vert\vartheta_1(\hat{x},\hat{u},\hat{w})\Vert \,\Vert x_1-x_l\Vert\\
       &+\Vert f(x_2,\hat{u},w_2)-f(x_{\bar l},\hat{u},w_{2}')\Vert +\Vert\vartheta_1(\hat{x},\hat{u},\hat w)\Vert \,\Vert x_2-x_{\bar l}\Vert\\
      &~~~~~~~\text{(using Lemma \ref{lema1_c} and \eqref{eq9a})}\\ 
       \le& 2\big(\Vert x_1-x_l\Vert +\Vert x_2-x_{\bar l}\Vert\big)\mathcal{L}_x(\hat u)+\big(\Vert w_1-w_1'\Vert +\Vert w_2-w_{2}'\Vert\big)\mathcal{L}_w(\hat u)\\
       \le&\;4\mathcal{L}_x(\hat u)\Vert\hat{\eta}_x\Vert+2\mathcal{L}_w(\hat u)\Vert\eta_w\Vert.
       \end{aligned}
       \end{equation*}
       The above inequality implies that  
       \begin{equation}
           \label{eq10_c}
           \beta(\vartheta^{*},x_1,x_2,w_1,w_2)\le \beta(\vartheta^{*},x_l,x_{\bar l},w_1',w_{2}')+\varrho,
       \end{equation}
       where $\varrho:=4\mathcal{L}_x(\hat u)\hat\eta_x+2\mathcal{L}_w(\hat u)\Vert{\eta}_w\Vert\mathbf{1}_{\mathrm{dim}(\mathcal{X})}$.
       Therefore, within any cell $\Phi_{\eta_x/2}(\hat{x})$, \eqref{eq10_c} implies that any optimal solution of SCP \eqref{c_scp} is always feasible for RCP \eqref{c_rcp}. In particular, any feasible solution of \eqref{c_scp} results in a growth bound $\kappa_\vartheta$ of the form \eqref{k_c_para} that satisfies inequality \eqref{kap_c_eqn}, which concludes the proof.
   \end{proof}

We propose Algorithm \ref{alg_sa} to illustrate the required procedures in Theorem \ref{c_unknown_data} for the data-driven construction of finite abstractions of subsystems. We present a visual representation of the transition function $\hat f(\hat x,\hat u,\hat w)$ for a finite abstraction of a subsystem in Fig. \ref{fig2}.

   \begin{algorithm}[ht!]
	\caption{Construction of Subsystems Abstractions (\emph{SA})}
 \label{alg_sa}
 \begin{algorithmic}[1]
\REQUIRE \emph{SA}\,(\,$\mathcal{X}$, $U$, $\mathcal{U}$, $\eta_x$, $\eta_u$, $\eta_w$) 
    \STATE{Construct: $\hat{\mathcal{X}}=[\mathcal{X}]_{\eta_x}$, $\hat U=[U]_{\eta_u}$, $\hat{\mathcal{U}}=[\mathcal{U}]_{\eta_w}$}
	\FOR{\textbf{each} $(\hat x,\hat u,\hat w)\in(\hat{\mathcal{X}},\hat U,\hat{\mathcal{U}})$}
		\STATE Initiate: $\hat f(\hat x,\hat u,\hat w)=\emptyset$, $\gamma = \textbf{0}$
  \STATE Obtain $c= f(\hat x,\hat u,\hat w)$ by simulating dt-CS \eqref{eq2.4} from the initial condition $\hat x$ under the input pair $(\hat u, \hat w)$ 
  \STATE Compute $\varrho\in\mathbb{R}^{\mathrm{dim}(\mathcal{X})}_{\ge0}$ as in \eqref{gamma_c}
  \STATE As outlined in Theorem \ref{c_unknown_data}, generate $[\Phi_{\eta_x/2}(\hat{x})]_{\hat\eta_x}$ and select $\mathcal{N}_C$ sampled data points $(x_l , \hat u,\hat w, x'_l )$ from it.
\STATE Obtain the optimal value $\vartheta^{*}(\hat x,\hat u,\hat w)$ for the SCP in \eqref{c_scp}
\STATE Update $\gamma=\kappa_{\vartheta^{*}}(\eta_x,\hat x,\hat u,\hat w)$
\STATE $\hat f (\hat x,\hat u,\hat w)=\{\hat x'\in\hat{\mathcal{X}}~|~\Phi_{\eta_x/2}( \hat x')\cap\Phi_{\gamma}(c)\neq\emptyset\}\cup\hat f (\hat x,\hat u,\hat w)$
\ENDFOR	
	\ENSURE $\widehat{\Xi}=(\hat{\mathcal{X}},\hat{U},\hat{\mathcal{U}},\hat{f})$
\end{algorithmic}
\end{algorithm}

\begin{remark}
    \label{rmk_subsys_scp}
    Utilizing Theorem \ref{c_unknown_data} across all partition sets enables the establishment of a set membership relation $\preceq_{\mathcal{Q}}$ (cf. Theorem \ref{sub_abs_thrm}) between a concrete subsystem $\Xi$ and its data-driven finite abstraction $\widehat\Xi$. This abstraction is constructed by solving SCP \eqref{c_scp} over the grid cells.
\end{remark}

\subsection{Finite Abstractions for Interconnections}
\label{sec4.2}
Upon obtaining the subsystem abstractions, as in the preceding subsection, we proceed to the second phase of addressing Problem \ref{prob1}. This phase involves creating a symbolic representation of the interconnection. The construction of arbitrary interconnections can, in the worst-case scenario, exhibit exponential complexity with respect to the number of subsystems $N$. To mitigate this, we introduce supplementary variables that account for intermediate computations within the abstraction construction. This not only introduces sparsity but also reduces the computational load associated with the abstraction process.

Consider an interconnection $\mathcal{I}=(\mathcal{X}^\mathcal{I},\mathcal{U}^\mathcal{I},\mathcal{M})$ as in Definition \ref{intercon}. Note that the interconnection map $\mathcal{M}$ \eqref{inter_map} is a partially unknown nonlinear map. However, we have access to the input-output data points as in $\mathcal{D}_{\mathcal{N}_I}$. Our goal is to derive a linear approximation $\widebar{\mathcal{M}}$ of $\mathcal{M}$, serving as the basis for constructing an abstraction $\hat{\mathcal{I}}$. To do so, we first frame the search for the linear estimate as a regression problem. The estimator is subject to regularization via an $\ell_1$-norm penalty term, which promotes sparsity. This is detailed as follows:
\begin{equation}
    \label{lasso}
    \argmin_{\widebar{\mathcal{M}}}\Big\{\big\Vert\Omega - \mathbb{X}\widebar{\mathcal{M}}\big\Vert_2^2+\alpha\big\Vert\widebar{\mathcal{M}}^\top\big\Vert_1\Big\},
\end{equation}
where for data points $\{(x_l,w_l)\}_{l=1}^{\Bar{\mathcal{N}}_I}$ in $\mathcal{D}_{\mathcal{N}_I}$, $\Omega$ and $\mathbb{X}$ are the stack of $\{w_l\}$ and $\{x_l\}$, respectively, with $l\in[1;\Bar{\mathcal{N}}_I]$ such that $\Bar{\mathcal{N}}_I<\mathcal{N}_I$. Note that it is not necessary to utilize all data points in $\mathcal{D}_{\mathcal{N}_I}$ to formulate equation \eqref{lasso} for estimating $\mathcal{M}$. Instead, we rely on only a subset of $\mathcal{D}_{\mathcal{N}_I}$. Note that for a network comprising of $N$ interconnected subsystems governed by a linear interconnection map, one can readily verify that acquiring $\mathrm{dim}(\mathcal{U}^\mathcal{I})\times \mathrm{dim}(\mathcal{X}^\mathcal{I})$ input-output data points of the interconnection map is sufficient for accurately estimating the map (cf. Subsection \ref{sec52}). Note that $\alpha>0$ denotes the regularization parameter and $\widebar{\mathcal{M}}\in\mathbb{R}^{\mathrm{dim}(\mathcal{U}^\mathcal{I})\times\mathrm{dim}(\mathcal{X}^\mathcal{I})}$. The term $\alpha\big\Vert\widebar{\mathcal{M}}^\top\big\Vert_1$ in \eqref{lasso} enforces sparsity on $\widebar{\mathcal{M}}$ row by row \cite{tibshirani1996regression}, thereby reducing the computational effort needed for constructing the overall abstraction. However, to accomplish the desired sparsity on $\widebar{\mathcal{M}}$ from \eqref{lasso}, an appropriate value must be selected for parameter $\alpha$. One may opt for the process of tuning the penalty term through cross-validation \cite{ljung1998system}, involving systematically assessing a range of penalty values on validation data. 

It is important to note that the regression formulation in \eqref{lasso} is numerically solvable, guaranteeing the existence of a solution for the specified optimization problem. Consequently, given $x_l\in\mathcal{X}^\mathcal{I}$, the actual value $w_l=\mathcal{M}(x_l)\in\mathcal{U}^\mathcal{I}$ is estimated by $\widebar{\mathcal{M}}x_l$, accounting for a residual error. We present the next lemma, which plays a crucial role in the process of constructing an abstraction for an interconnection $\mathcal{I}$.

\begin{lemma}
    \label{lem1}
    Consider a data point $(x_l,w_l)\in\mathcal{D}_{\mathcal{N}_I}$ and a neighbourhood $\Phi_{\rho_x}(x_l)\subset\mathcal{X}^\mathcal{I}$ of $x_l$, where $\rho_x\in\mathbb{R}_{>0}^{\mathrm{dim}(\mathcal{X}^\mathcal{I})}$. 
    Then, for any $x'\in\Phi_{\rho_x}(x_l)$, there exists $\hat\varepsilon(\rho_x,w_l,x')\in\mathbb{R}_{>0}$ such that \\$\mathcal{M}(x')\in\widebar{\mathcal{M}}x'\oplus[-\hat\varepsilon(\rho_x,w_l,x')\mathbf{1}_{\mathrm{dim}(\mathcal{U}^\mathcal{I})},\,\hat\varepsilon(\rho_x,w_l,x')\mathbf{1}_{\mathrm{dim}(\mathcal{U}^\mathcal{I})}]$.
\end{lemma}
\begin{proof}
    Consider any $x'\in\mathcal{X}^\mathcal{I}$ where $|x_l-x'|\leq\rho_x$. Then the following inequalities hold:
    \begin{equation*}
    \begin{aligned}
        \Vert\mathcal{M}(x')-\widebar{\mathcal{M}}x'\Vert&\leq\Vert\mathcal{M}(x')-\mathcal{M}(x_l)\Vert+\Vert\mathcal{M}(x_l)-\widebar{\mathcal{M}}x'\Vert\\
        &\leq\mathcal{L}_\mathcal{M}\Vert x'-x_l\Vert+\Vert w_l-\widebar{\mathcal{M}}x'\Vert\\
&\leq\mathcal{L}_\mathcal{M}\Vert\rho_x\Vert+\Vert w_l-\widebar{\mathcal{M}}x'\Vert.
        \end{aligned}
    \end{equation*}
    Now, let
    \begin{equation}
        \label{eps_hat}
\hat\varepsilon(\rho_x,w_l,x'):=\mathcal{L}_\mathcal{M}\Vert\rho_x\Vert+\Vert w_l-\widebar{\mathcal{M}}x'\Vert. 
    \end{equation}
Therefore, for any $x'\in\Phi_{\rho_x}(x_l)$, $\Vert\mathcal{M}(x')-\widebar{\mathcal{M}}x'\Vert\leq\hat\varepsilon(\rho_x,w_l,x')$, which concludes the proof.
\end{proof}

Observe that the parameter $\hat\varepsilon(\rho_x,w_l,x')$ depends on the data point $(x_l,w_l)\in\mathcal{D}_{\mathcal{N}_I}$, along with the distance $\rho_x\in\mathbb{R}_{>0}^{\mathrm{dim}(\mathcal{X}^\mathcal{I})}$ of the neighborhood around $x_l$, considering the particular point $x'$ within the neighborhood. Using $\widebar{\mathcal{M}}$ to derive an over-approximation of $\mathcal{M}(x')$ in accordance with Lemma \ref{lem1} implies creating an interconnection abstraction in a monolithic fashion. This approach  necessitates a brute-force traversal of the set $\mathcal{X}^\mathcal{I}$. The computational complexity of this brute-force exploration experiences an exponential increase with the number of subsystems $N$. Hence, we illustrate how integrating additional variables, representing intermediate computations through the decomposition of $\widebar{\mathcal{M}}$, can reduce the runtime of the interconnection abstraction. This reduction occurs because the brute-force traversal is conducted over lower-dimensional subspaces, consequently alleviating the computational burden associated with the abstraction process.

Consider a $x\in\Phi_{\rho_x}(x_l)$ according to Lemma \ref{lem1}. Let
\begin{equation}
    \label{dec}
w=\widebar{\mathcal{M}}x+\hat\varepsilon(\rho_x,w_l,x)\mathbf{1}_{\mathrm{dim}(\mathcal{U}^\mathcal{I})}.
\end{equation}
We proceed by transforming the sparse matrix $\widebar{\mathcal{M}}\in\mathbb{R}^{\mathrm{dim}(\mathcal{U}^\mathcal{I})\times\mathrm{dim}(\mathcal{X}^\mathcal{I})}$ into a weighted directed acyclic graph (WDAG) $G=(V,E)$ such that the following hold:
\begin{itemize}
    \item The sets of vertices $w_I=\{w_1,\ldots,w_N\}$ and $x_I=\{x_{1},\ldots,x_{N}\}$ satisfy $w_I\cup x_I\subseteq V$.
    \item The edge function $e:x_I\times w_I\rightarrow\mathbb{R}\cup\{\text{invalid}\}$ is defined as follows: 
    \begin{equation}
   \label{e_func}
   e(x_{i},w_j):=
   \begin{cases}
       \text{invalid }&\text{if }\widebar{\mathcal{M}}_{ij}=0\\
       \widebar{\mathcal{M}}_{ij}&\text{ otherwise.}
   \end{cases}
   \end{equation}
   \item The set of edges $E=\{(v,v')\in V\times V~|~e(v,v')\text{ is not invalid}\}$. 
\end{itemize}
We emphasize that an edge $(v,v')\in V\times V$ is considered invalid if it is not included in the graph $G$, meaning $(v,v')\notin E$. Additionally, for any vertex $v\in V$, we define the set $\mathcal{D}(v)\!:=\!\{v'~|~e(v',v)\text{ is not invalid}\}$. Consequently, the indegree of $v$, denoted as $deg(v)$, is defined as $\mathcal{C}_d(\mathcal{D}(v))$. We proceed to decompose the matrix $\widebar{\mathcal{M}}$ by leveraging its corresponding WDAG, applying the procedures outlined in Algorithm \ref{alg_dec}. It is important to mention that Algorithm \ref{alg_dec} takes a user-specified hyperparameter $\sigma\in\mathbb{N}_{\ge 1}$ as an input, which gives an upper bound on the indegree of all vertices in the WDAG and determines the number of nodes in its final layer. To mitigate potential computational challenges (see Lines \( 4 \) to \( 8 \) of Algorithm \ref{alg_ia}), it is advisable to limit \( \sigma \) to a maximum value of 4. This recommendation stems from the limitations of current software tools (\emph{e.g.,} \cite{mouelhi2013cosyma, mazo2010pessoa, rungger2016scots}) designed for abstracting continuous-space control systems into finite-state abstract models, which are generally restricted to modestly sized systems (up to 4 dimensions). These tools experience exponential runtime growth with respect to the dimensions of the state and input sets, making the abstraction of larger systems computationally infeasible.

\begin{figure}[ht]
\centering   
\subfigure{\label{declf}\includegraphics[width=60mm]{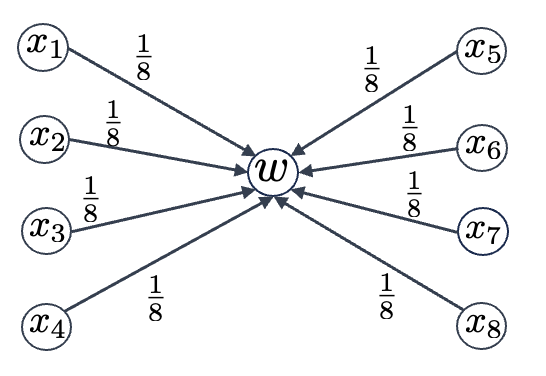}}
\subfigure{\label{decrght}\includegraphics[width=80mm]{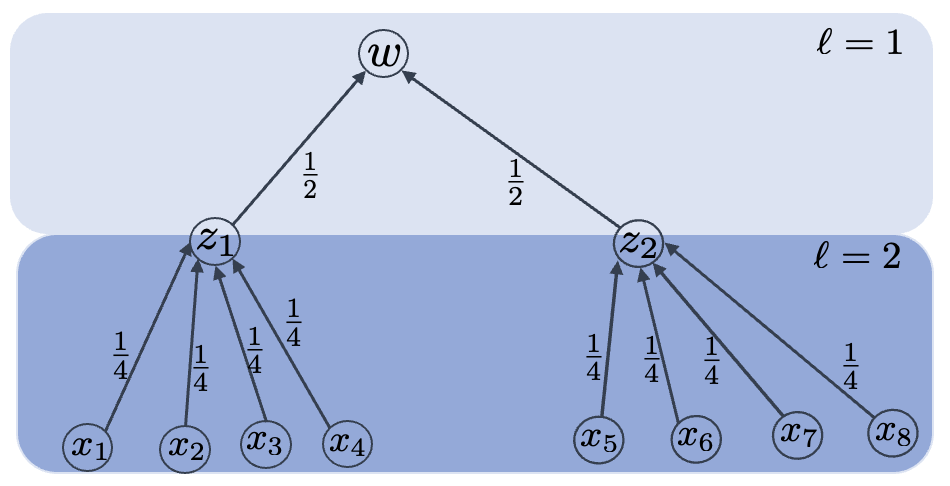}}
\caption{An illustration of a WDAG with the application of Algorithm \ref{alg_dec} where $\sigma=4$.}
\label{figdec}
\end{figure}

In Fig. \ref{figdec} (top), we provide an example of a WDAG of $\widebar{\mathcal{M}}\in\mathbb{R}^{1\times8}_{>0}$ with a uniform weight of $1/8$ on each edge. This example is further explained in Subsection \ref{sec52}. The graph features a shared node $w$ such that $deg(w)=8$. Applying Algorithm \ref{alg_dec} with $\sigma=4$ yields the lower part of the figure. Two intermediate variables $\{z_1,z_2\}$ have been introduced to the graph, effectively decomposing $\widebar{\mathcal{M}}\in\mathbb{R}^{1\times8}_{>0}$ to $M_1\in\mathbb{R}^{1\times2}_{>0}$ and $M_2\in\mathbb{R}^{1\times4}_{>0}$. 

In the abstraction process, instead of over-approximating $\widebar{\mathcal{M}}\hat x$ for a $\hat x\in\mathcal{X}^\mathcal{I}\subset \mathbb{R}^{8}_{\ge0}$, we leverage the decomposition of $\widebar{\mathcal{M}}$ to compute the over-approximation. Thus, in this example, \eqref{dec} which involves computations in $\mathbb{R}^8$ is decomposed into computations in $\mathbb{R}^2$ and $\mathbb{R}^4$. Consequently, the set $\widebar{\mathcal{M}}\hat x\oplus[-\hat\varepsilon(\rho_x,w_l,\hat x),\,\hat\varepsilon(\rho_x,w_l,\hat x)]$ is over-approximated as $\{\prod_{i=1}^2(M_ix_i\oplus[-\hat\varepsilon(\rho_x,w_l,\hat x),\,\hat\varepsilon(\rho_x,w_l,\hat x)])~|~\hat x=[x_1;x_2]\}$
, and grids of dimension $2$ and $4$ are availed instead of dimension $8$ for the abstraction. 

\begin{algorithm}[!ht]
	\caption{Matrix Decomposition (\emph{MDec})}
 \label{alg_dec}
 \begin{algorithmic}[1]
\REQUIRE \emph{MDec}\,\big($\widebar{\mathcal{M}}$,\,$\sigma$\big)
    \STATE Generate the corresponding WDAG $G=(V=w_I\cup x_I,E)$ with weights on the edge according to \eqref{e_func}
    \STATE Initiate matrix set $MS=\emptyset$ 
    \WHILE{$deg(v)>\sigma~\forall v\in V$}
    \STATE Initiate set of auxilliary variables $Z_\ell=\emptyset$ with layer $\ell=1$
    \STATE Pick $d\in[1;\sigma]$ and create vertices $z_i$ where $1\le i\le d$
    \STATE Let $p:=\mathcal{C}_d(w_I)$ and create matrix $M_\ell\in\mathbb{R}^{p\times d}$ with entries $e(z_i,v)$ set to null $\forall v\in w_I$ and $1\le i\le d$
	\FOR{\textbf{each} $v\in w_I$}
 \IF{$deg(v)\le \sigma$}
 \STATE Continue
 \ELSE 
 \STATE Assign weight value $1/d$ to edge $(z_i,v)$
 \STATE Compute $e(z_i,v)$ according to \eqref{e_func} to update matrix $M_\ell$ and let $d':=\mathcal{C}_d(\mathcal{D}(v))$ 
 \STATE $D:=[v_1';v_2';\ldots;v'_{d'}]$ where $v_j'\in\mathcal{D}(v)~\forall j\in[1;d']$
 \STATE Set $e(D[j],z_i):=e(D[j],v)/e(z_i,v)$ until all vertices $D[j]$ has been connected to a vertex $z_i$ where \\$1\le j\le d'$ and $1\le i\le d$
 \STATE $E=E\setminus\{(v',v)\in V\times V~|~v'\in\mathcal{D}(v)\}$
 \STATE $Z_\ell=Z_\ell\cup\{z_i~|~1\le i\le d\}$
 \ENDIF
	\ENDFOR
 \STATE $w_I=Z_\ell$, $\ell=\ell+1$ and $MS=MS\cup\{M_\ell\}$ 
 \ENDWHILE
	\ENSURE $MS$
\end{algorithmic}
\end{algorithm}

Furthermore, we propose Algorithm \ref{alg_ia} to outline the process of constructing the interconnection abstraction. Note that in line $5$ of Algorithm \ref{alg_ia}, the computational burden associated with traversing a higher-dimensional space $\mathcal{X}^{\hat{\mathcal{I}}}$ is mitigated through the decomposition of $\widebar{\mathcal{M}}$ in line $4$. This decomposition is utilized in the construction of the over-approximation, as outlined in line $8$.
\begin{algorithm}[!ht]
	\caption{Construction of Interconnection Abstraction (\emph{IA})}
 \label{alg_ia}
 \begin{algorithmic}[1]
\REQUIRE \emph{IA}\,(\,$\mathcal{X}^\mathcal{I}$, $\mathcal{U}^\mathcal{I}$, $\rho_x$, $\rho_w$)
    \STATE{Initiate: 
    $\mathcal{X}^{\hat{\mathcal{I}}}=[\mathcal{X}^\mathcal{I}]_{\rho_x}$ and $\mathcal{U}^{\hat{\mathcal{I}}}=[\mathcal{U}^\mathcal{I}]_{\rho_w}$}
    \STATE{Collect data in set $\mathcal{D}_{\mathcal{N}_I}=\{(x_l,w_l)\in\mathcal{X}^\mathcal{I}\times\mathcal{U}^\mathcal{I}~|~x_l\in\Phi_{\rho_x}(\hat x)\text{ where }\hat x\in\mathcal{X}^{\hat{\mathcal{I}}},\;l=1,\ldots,\mathcal{N}_I\}$ and select appropriate value for $\sigma$ } 
    \STATE{Using $\Bar{\mathcal{N}}_I<\mathcal{N}_I$ data points, solve \eqref{lasso} to obtain $\widebar{\mathcal{M}}$}
    \STATE Obtain $MS:=MDec(\widebar{\mathcal{M}},\,\sigma)$ and let $q:=\mathcal{C}_d(MS)$ 
	\FOR{\textbf{each} $\hat x\in\mathcal{X}^{\hat{\mathcal{I}}}$}
		\STATE Initiate $\hat{\mathcal{M}}(\hat x)=\emptyset$ and compute $\hat\varepsilon:=\hat\varepsilon(\rho_x,w_l,\hat x)>0$ as in \eqref{eps_hat}
  \STATE Construct lower dimensional grids $[\mathcal{X}^\mathcal{I}]^i_{\Vert\rho_x\Vert\mathbf{1}_i}$ where $i\in[1;q]$
\STATE $B\!:=\!\{\prod_{i=1}^q(M_ix_i\oplus[-\hat\varepsilon\mathbf{1}_i,\,\hat\varepsilon\mathbf{1}_i])~|~\hat x=[x_1;\ldots ;x_q]\}$
\STATE $\hat{\mathcal{M}}(\hat x)=\{\hat w'\in\mathcal{U}^{\hat{\mathcal{I}}}~|~\Phi_{\rho_w}( \hat w')\cap B\neq\emptyset\}\cup\hat{\mathcal{M}}(\hat x)$
\ENDFOR	
	\ENSURE $\hat{\mathcal{I}}=(\mathcal{X}^{\hat{\mathcal{I}}},\mathcal{U}^{\hat{\mathcal{I}}},\hat{\mathcal{M}})$
\end{algorithmic}
\end{algorithm}
Additionally, we formally establish that the interconnection created by using Algorithm \ref{alg_ia} is a correct abstraction of the concrete \emph{partially unknown} interconnection. 
\begin{theorem}
    \label{thrm_inter}
    Consider an interconnection $\mathcal{I}=(\mathcal{X}^{\mathcal{I}},\mathcal{U}^{\mathcal{I}},\mathcal{M})$ where the map $\mathcal{M}$ is partially unknown. Let $\rho_x\in\mathbb{R}_{>0}^{\mathrm{dim}(\mathcal{X}^\mathcal{I})}$ and $\rho_w\in\mathbb{R}_{>0}^{\mathrm{dim}(\mathcal{U}^\mathcal{I})}$ be discretization parameters utilized to form symbolic sets $\mathcal{X}^{\hat{\mathcal{I}}}=[\mathcal{X}^\mathcal{I}]_{\rho_x}$ and $\mathcal{U}^{\hat{\mathcal{I}}}=[\mathcal{U}^\mathcal{I}]_{\rho_w}$. Suppose $\hat{\mathcal{I}}=(\mathcal{X}^{\hat{\mathcal{I}}},\mathcal{U}^{\hat{\mathcal{I}}},\hat{\mathcal{M}})$ is constructed using Algorithm \ref{alg_ia}.  
    Then $\hat{\mathcal{I}}$ is an abstraction of $\mathcal{I}$ as in Definition \ref{intercon}.
\end{theorem}
\begin{proof}
    Suppose $(x,\hat x)\in\mathcal{X}^\mathcal{I}\times\mathcal{X}^{\hat{\mathcal{I}}}$ such that $|x-\hat x|\leq\rho_x$ and $w=\mathcal{M}(x)$. Line $1$ of Algorithm \ref{alg_ia} implies that there exists $\hat w\in\mathcal{U}^{\hat{\mathcal{I}}}$ such that $w\in\Phi_{\rho_w}(\hat w)$. Upon solving \eqref{lasso}, we obtain an approximation $\widebar{\mathcal{M}}$ of $\mathcal{M}$. Then Lemma \ref{lem1} implies that there exists $\hat\varepsilon=\hat\varepsilon(\rho_x,w_l,\hat x)>0$ such that $|\mathcal{M}(x)-\widebar{\mathcal{M}}x|\leq\hat\varepsilon\mathbf{1}_{\mathrm{dim}(\mathcal{U}^\mathcal{I})}$. Hence, $w\in\Phi_{\hat\varepsilon\mathbf{1}_{\mathrm{dim}(\mathcal{U}^\mathcal{I})}}(\widebar{\mathcal{M}}x)$, which implies that $\Phi_{\rho_w}(\hat w)\cap\Phi_{\hat\varepsilon\mathbf{1}_{\mathrm{dim}(\mathcal{U}^\mathcal{I})}}(\widebar{\mathcal{M}}x)\neq\emptyset$. Therefore, line $9$ of Algorithm \ref{alg_ia} yields that $\hat w\in\hat{\mathcal{M}}(\hat x)$, which satisfies the condition outlined in Definition \ref{inter_abs}, concluding the proof.
\end{proof}

\subsection{Compositional Construction of Abstraction}
Here, we begin by presenting a compositional result that establishes a feedback refinement relation between an interconnected abstraction and the concrete interconnected dt-CS. This result leverages the feedback refinement relations from individual concrete subsystems to their respective abstractions. Subsequently, we outline a scheme for integrating the subsystem abstraction and the interconnection abstraction in a compositional manner. 
Suppose we have $N$ subsystems $\Xi_i=(\mathcal{X}_i,U_i,\mathcal{U}_i,f_i)$
, accompanied by their finite abstractions $\widehat{\Xi}_i=(\hat{\mathcal{X}}_i,\hat{U}_i,\hat{\mathcal{U}}_i,\hat{f}_i)$
, and feedback refinement relations $\mathcal{Q}_i$ from $\Xi_i$ to $\widehat{\Xi}_i$, where $i\in[1;N]$ (cf. Theorem \ref{sub_abs_thrm}). 
We present the following theorem, which establishes a feedback refinement relation between a concrete interconnected dt-CS and the data-driven finite abstractions of the subsystems that have been composed together.
\begin{theorem}
    \label{asf_thrm}
    Consider an interconnected dt-CS $\Xi\!=\!\mathcal{I}(\Xi_1,\ldots,\Xi_N)$, consisting of $N\in\mathbb{N}_{\ge 1}$ subsystems $\Xi_i$ and an interconnection $\mathcal{I}=(\mathcal{X}^\mathcal{I},\mathcal{U}^\mathcal{I},\mathcal{M})$. Let each subsystem $\Xi_i$ admits a data-driven abstraction $\widehat\Xi_i$ constructed as in Algorithm \ref{alg_sa} such that $\Xi_i\preceq_{\mathcal{Q}_i}\widehat{\Xi}_i$ $\forall i\in[1;N]$ (\emph{e.g.} as in Theorem \ref{sub_abs_thrm}). Additionally, assume that $\hat{\mathcal{I}}=(\mathcal{X}^{\hat{\mathcal{I}}},\mathcal{U}^{\hat{\mathcal{I}}},\hat{\mathcal{M}})$ is an abstraction of 
    $\mathcal{I}$ constructed as in Algorithm \ref{alg_ia}. Consider a relation $Q\subseteq\mathcal{X}^{\mathcal{I}}\times\mathcal{X}^{\hat{\mathcal{I}}}$ such that $\forall(x,\hat x)\in Q$ where $x=[x_1;\ldots;x_N]$ and $\hat x=[\hat{x}_1;\ldots;\hat{x}_N]$, $(x_i,\hat{x}_i)\in\mathcal{Q}_i$ $\forall i\in[1;N]$. Then relation $Q$ is a feedback refinement relation from $\Xi$ to $\hat{\mathcal{I}}(\widehat\Xi_1,\ldots,\widehat\Xi_N)$.
\end{theorem}
\begin{proof}
    \label{frr_proof}
    Given that $\Xi_i\preceq_{\mathcal{Q}_i}\widehat{\Xi}_i$, $\forall i\in[1;N]$, let $\hat u:=[\hat{u}_1;\ldots;\hat{u}_N]\in U_{\hat{\mathcal{I}}(\widehat\Xi_1,\ldots,\widehat\Xi_N)}(\hat x)$ where $\hat x=[\hat{x}_1;\ldots;\hat{x}_N]\in\mathcal{X}^{\hat{\mathcal{I}}}$. Thus, $\exists\hat{x}'\in\hat{f}(\hat x,\hat u)$, and by \eqref{sub_f} $\hat f(\hat x, \hat u):=\{[\hat {x}_1';\ldots;\hat {x}_N']~|~\hat{x}_i'\in\hat{f}_i(\hat{x}_i,\hat{u}_i,\hat{w}_i)~\forall i\in[1;N]\}$ where $\forall w:=[w_1;\ldots;w_N]=\mathcal{M}([x_1;\ldots;x_N])$ $\exists\hat w:=[\hat{w}_1;\ldots;\hat{w}_N]\in\hat{\mathcal{M}}(\hat x)$ and there is a relation $\tilde{\mathcal{R}}\subseteq\mathcal{U}^\mathcal{I}\times\mathcal{U}^{\hat{\mathcal{I}}}$ such that $(w,\hat w)\in\tilde{\mathcal{R}}$. Suppose that $(x,\hat x)\in Q$ where $x=[x_1;\ldots;x_N]\in\mathcal{X}^\mathcal{I}$ and $(x_i,\hat x_i)\in\mathcal{Q}_i$, $\forall i\in[1;N]$. Thus, $\hat u_i\in U_{\hat\Xi_i}(\hat x_i,\hat w_i)\implies \mathcal{Q}_i(f_i(x_i,\hat u_i,w_i))\subseteq\hat{f}(\hat x_i,\hat u_i,\hat w_i)$ $\forall i\in[1;N]$. Since $\hat x_i\in\hat{f}(\hat x_i,\hat u_i,\hat w_i)$, $\forall i\in[1;N]$, and by Definition \ref{asf}, $\exists w_i\in\mathcal{U}_i$ and $x_i'\in\mathcal{X}_i$ such that $x_i'= f_i(x_i,\hat u_i,w_i)$ $\forall i\in[1;N]$. Therefore, $x'= f(x, \hat u):=\{[x_1';\ldots;x_N']~|~x_i'=f_i(x_i,\hat u_i,w_i)~\forall i\in[1;N]\}$, and as a result, $\hat u\in U_{\Xi}(x)$, which implies that $U_{\hat{\mathcal{I}}(\widehat\Xi_1,\ldots,\widehat\Xi_N)}(\hat x)\subseteq U_{\Xi}(x)$. Now, suppose that $\hat x'\in Q(f(x,\hat u))$, then $\hat x'\in \mathcal{Q}_i(f_i(x_i,\hat u_i,w_i))$ $\forall i\in[1;N]$. This implies $\hat x_i'\in\hat{f}_i(\hat x_i,\hat u_i,\hat w_i)$, $\forall i\in[1;N]$, which implies that $\hat x'\in \hat f(\hat x,\hat u)$. Hence, $Q(f(x,u))\subseteq\hat{f}(\hat x,\hat u)$, and the conditions in Definition \ref{asf_mono} are satisfied, concluding the proof.
\end{proof}

\begin{remark}
    \label{rmk3.2.1}
    Theorem \ref{asf_thrm} enables the construction of a finite abstraction of a network of subsystems by utilizing finite abstractions of individual subsystems, as depicted in Fig. \ref{fig1}. 
\end{remark} 

The abstractions derived for both the subsystems and the interconnection via Algorithm \ref{alg_sa} and \ref{alg_ia}, respectively, are encoded using binary decision diagrams (BDD) \cite{bryant1986graph}. This selection of the data structure efficiently encodes the abstractions as Boolean-valued functions. Hence, each subsystem abstraction have an encoding $(\hat x_i,\hat u_i,\hat w_i, \hat x_i')-{\hat f_i}\rightarrow\{\texttt{true},\texttt{false}\}$, $\forall i\in[1;N]$. Similarly, the interconnection abstraction is encoded as $(\hat x,\hat w')-{\hat{\mathcal{M}}}\rightarrow\{\texttt{true},\texttt{false}\}$. Consequently, we utilize this representation of abstractions to combine and interconnect the subsystem abstractions through Boolean conjunctions, taking into account all conditions as delineated in Definition \ref{intercon2}. Thus, we outline the complete solution to Problem \ref{prob1} in Algorithm \ref{alg_ca}.

\begin{algorithm}[ht!]
	\caption{Data-Driven Compositional Abstraction (\emph{CA})}
 \label{alg_ca}
 \begin{algorithmic}[1]
\REQUIRE \emph{CA}\,\big(\,$\mathcal{I}(\Xi_1,\ldots,\Xi_N)$, $\{\eta_{x_i},\eta_{u_i},\eta_{w_i}\}_{i=1}^N$, $\rho_x$, $\rho_w$\big)
    \STATE Initiate $Abs=\texttt{true}$
	\FOR{\textbf{each} $i\in[1;N]$}
		\STATE $Abs'=$ \emph{SA}\,(\,$\Xi_i,\eta_{x_i},\eta_{u_i},\eta_{w_i}$)
  \STATE $Abs=Abs\land Abs'$
	\ENDFOR
 \STATE $Abs=Abs\;\land$ \emph{IA}\,\big(\,$\mathcal{I}(\Xi_1,\ldots,\Xi_N)$, $\rho_x$, $\rho_w$\big) 
	\ENSURE $Abs$
\end{algorithmic}
\end{algorithm}

Constructing the abstraction of the interconnected system monolithically, as done in the literature (\emph{e.g.,} \cite{ajeleye2023data, kazemi2022data,gruber2017sparsity}), results in a brute-force algorithm with complexity
$\mathcal{O}\big(\prod_{i=1}^N \mathcal{C}_d([\mathcal{X}_i]_{\eta_{x_i}})\,\mathcal{C}_d([U_i]_{\eta_{u_i}})\big)$. In contrast, our proposed compositional approach (\emph{i.e.,} Algorithm \ref{alg_ca}) achieves a reduced complexity of \( \mathcal{O}\big(g^\sigma + \mathrm{dim}(\mathcal{X}_i)^{N}\mathrm{dim}(\mathcal{U}_i)^N\sigma + \sum_{i=1}^N \mathcal{C}_d([\mathcal{X}_i]_{\eta_{x_i}})\,\mathcal{C}_d([U_i]_{\eta_{u_i}})\, \mathcal{C}_d([\mathcal{U}_i]_{\eta_{w_i}}) \big) \), where \( g = \mathcal{C}_d([\mathcal{X}^\mathcal{I}]_{\rho_x}) \) and \( \sigma \) is the maximum in-degree of all vertices in the WDAG from Algorithm \ref{alg_dec}. By setting the user-specified hyperparameter \( \sigma \leq 4 \) in Algorithm \ref{alg_dec}, the compositional approach offers significantly lower computational complexity as the dimensions of the state set, input set, and number of subsystems increase, compared to monolithic methods. Consequently, the monolithic approach's complexity grows exponentially with the number of subsystems, resulting in a significantly larger abstraction size. In contrast, our proposed compositional approach maintains linear complexity relative to the number of subsystems.

\begin{remark}
Our proposed compositional approach for constructing finite abstractions of interconnected systems relies on the Lipschitz continuity of subsystem transition functions and the interconnection map, limiting its applicability to such systems. By leveraging subsystem interconnections, the approach reduces high-dimensional computations to lower dimensions, shifting sample complexity to the subsystem level. When the interconnection map is fully unknown but linear, the estimation procedure in \eqref{lasso} accurately identifies the map \( \mathcal{M} \) as the matrix \( \widebar{\mathcal{M}} \) using a minimum of \( \mathrm{dim}(\mathcal{U}^\mathcal{I}) \times \mathrm{dim}(\mathcal{X}^\mathcal{I}) \) samples, allowing the Lipschitz constant \( \mathcal{L}_{\mathcal{M}} \) to be estimated as \( \big\Vert \widebar{\mathcal{M}} \big\Vert \). For fully unknown nonlinear maps, Appendix \ref{apdx1} outlines an asymptotic estimation method, though quantitative estimates with confidence levels may also be used. Sample-efficient methods such as those in \cite{huang2023sample, knuth2021planning} could further improve the estimation process.
\end{remark}

\begin{remark}
    Our proposed compositional approach may face increased computational complexity when the interconnection map is fully connected, as the WDAG in Algorithm \ref{alg_dec} becomes fully connected, making it difficult to select a low value such as \( \sigma \leq 4 \). While a sparse linear approximation \( \widebar{\mathcal{M}} \) of the interconnection map \( \mathcal{M} \) can be constructed, the error introduced in Lemma \ref{lem1} may become significant if \( \mathcal{M} \) lacks sufficient sparsity, potentially leading to an overly conservative system abstraction.
\end{remark}

\section{Case Studies} \label{sec5}
In this section, we demonstrate the efficacy of our proposed approaches by applying them to three numerical benchmark models. Two of these are drawn from \cite{kim2018constructing}, while the third is an extended version of the tank model adapted from \cite{ajeleye2024data}. However, we have assumed that these models and their interconnection maps are partially unknown in applying our approach. The abstractions for both subsystems and the interconnection are symbolically represented using Boolean functions, which have been implemented with the \texttt{CUDD} toolbox \cite{somenzi1997cudd}. All implementations for the construction of the data-driven finite abstractions have been carried out in a modified version of the toolbox \texttt{SCOTS} \cite{rungger2016scots}, using a $64$-bit MacBook Pro with $64$GB RAM ($3.2$ GHz).

 \subsection{Experimental Evaluation for Scalability}
 \label{sec51}
We now present an example designed to illustrate the scalability of the interconnection decomposition introduced in Algorithm \ref{alg_dec} for
a range of $N$ subsystems. Consider a collection of $N$ scalar subsystems $\Xi_i=(\mathcal{X}_{i},U_{i},\mathcal{U},f_{i})$ 
where $\mathcal{X}_i=[0,32]$, $U_i=[0,7]$, $\mathcal{U}=[0,32]$ and 
\begin{equation}
    \label{ex1_f}
    \begin{split}
        x_i(k+1)&=f_i(x_i(k),u_i(k),w(k))\\
        &=\min(0.75(x_i(k)+u_i(k)),w(k)+1,32),
    \end{split}
\end{equation}
such that there is a single variable $w(k)\in\mathcal{U}$ shared amongst all subsystems, \emph{i.e.,}
\begin{equation}
    \label{ex1_w}
    w(k)=\max(x_1(k),\ldots,x_N(k)).
\end{equation}
It is evident from models in \eqref{ex1_f} and \eqref{ex1_w} that a one-time-step evolution of any subsystem depends on the previous state of other subsystems. Utilizing these models, we generate the data sets $\mathcal{D}_{\mathcal{N}_{C_i}}$ and $\mathcal{D}_{\mathcal{N}_I}$, respectively. The abstract sets $\hat{\mathcal{X}}_i$ and $\hat{\mathcal{U}}$ were constructed using grid parameters $\eta_{x_i}=\eta_{w}=1$ for all $i\in[1;N]$, while the external input grid was defined as $\hat U_i=\left[U_i\right]_1$. Furthermore, for all $\hat u\in \hat U_i$, we estimate the Lipschitz constants as $\mathcal{L}_{x_i}(\hat u)=\mathcal{L}_{w_i}(\hat u)=\mathcal{L}_{\mathcal{M}}=1.0$ (cf. Remark \ref{rmk_assump} and \cite[Lemma A.2.]{salamati2024data}). In Algorithm \ref{alg_sa}, we employed $\mathcal{N}_{C_i}=150$ samples. To construct the interconnection abstraction $\hat{\mathcal{I}}=\big(\prod^N_{i=1}\hat{\mathcal{X}}_i,\hat{\mathcal{U}},\hat{\mathcal{M}}\big)$, we solved \eqref{lasso} using $\Bar{\mathcal{N}}_I=200$ input-output data points from \eqref{ex1_w}. After obtaining $\widebar{\mathcal{M}}\in\mathbb{R}^N$, a decomposition $M_1:=\frac{1}{N}\widebar{\mathcal{M}}\cdot[\mathbf{1}_{N}~\;\mathbf{1}_{N}]$ follows from Algorithm \ref{alg_dec}  
to introduce intermediate variables 
as follows:
\begin{equation}
    \begin{split}
        z_1&=M_1\cdot[x_1;x_2]^\top,\\
        z_i&=M_1\cdot[z_{i-1};x_{i+1}]^\top,~\;\forall i\in[2;N-2],\\
        w&=M_1\cdot[z_{N-2};x_N]^\top,
    \end{split}
\end{equation}
where the intermediate variables $z_j,~\forall j\in[1;N-2]$, belongs to $\hat{\mathcal{X}}_i$.
\begin{figure}[!ht]
    \centering
    \includegraphics[width=12cm]{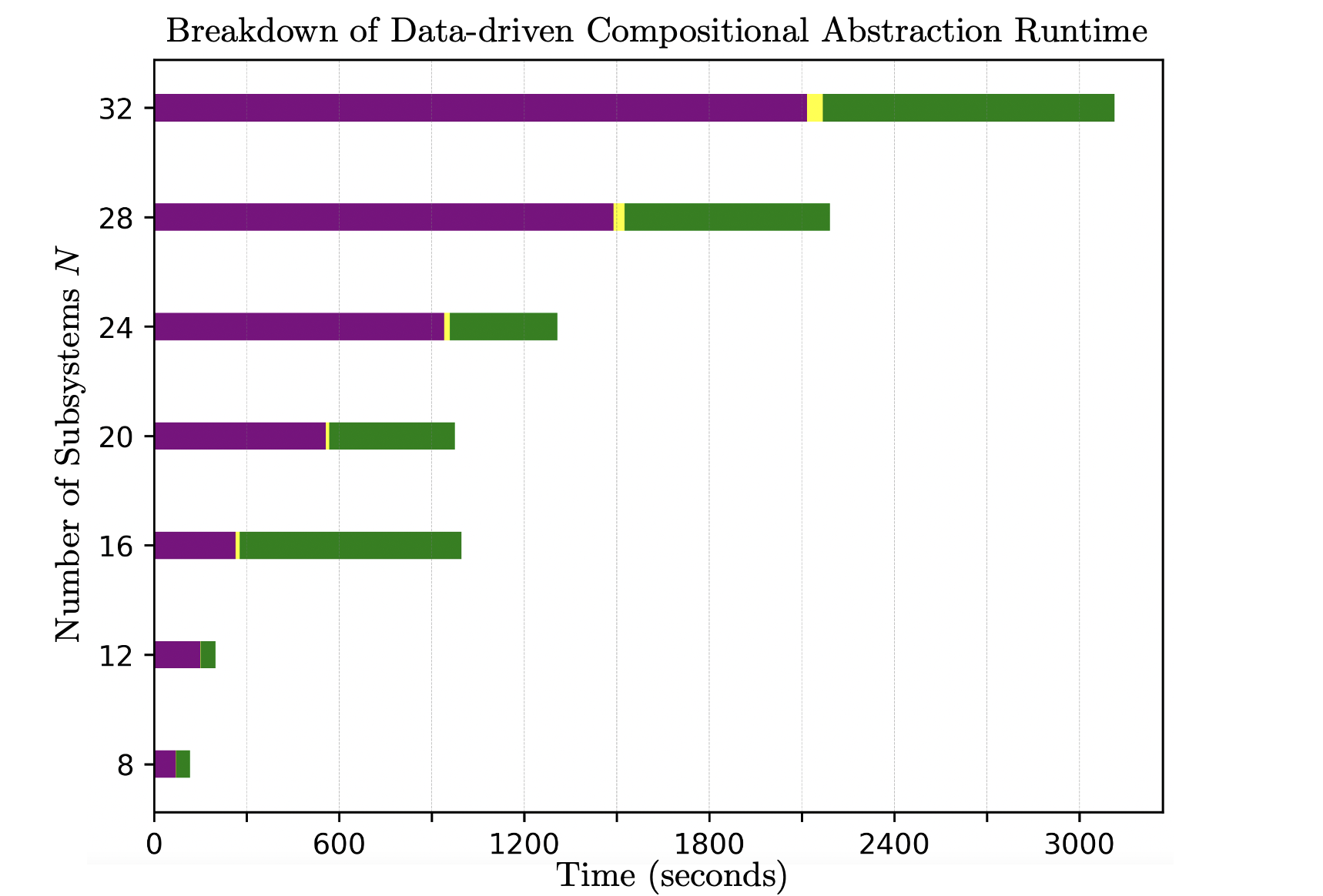}
    \caption{The runtime in seconds for systems in \eqref{ex1_f} and interconnection in \eqref{ex1_w} are shown for varying $N$. The cumulative runtime is represented by the horizontal bar, segmented into three colors: purple (left) represents abstracting and stacking of the abstraction BDD representations for all subsystems, yellow (middle) denotes the time for approximating the interconnection, and green (right) indicates composing the subsystems and interconnection abstractions.}
    \label{fig3}
\end{figure}
Fig. \ref{fig3} reports the runtimes required to construct both the $N$ abstract subsystems and the abstract interconnection, as well as their composition.
\subsection{Consensus-Driven Network of Logistic Systems} \label{sec52}
 We consider a network consisting of $N$ subsystems, each characterized by state sets $\mathcal{X}_i=[0,32]$ and input sets $U_i=[-2,2]$, where $i\in[1;N]$. Additionally, the subsystems share an internal input $\mathcal{U}=[0,32]$. The dynamics governing each subsystem $\Xi_i$ are defined as follows:
 \begin{equation}
 \label{ex2_f}
     \begin{split}
         x_i(k+1)&=f_i(x_i(k),u_i(k),w_i(k))\\
         &=\texttt{GLOG}_{[a,b]}(x_i(k)+u_i(k)+0.1(x_i(k)-w(k))),
     \end{split}
 \end{equation}
where $\texttt{GLOG}_{[a,b]}(\cdot)$ represents a generalized logistic function over $[a,b]\subseteq[0,32]$, producing output values within the range $[0,32]$. Mathematically, this function is defined as:
\begin{equation*}
    \texttt{GLOG}_{[a,b]}(x):=\frac{32}{1+e^{-0.2(x-\frac{b+a}{2})}}.
\end{equation*}
The function 
$\texttt{GLOG}_{[a,b]}(\cdot)$ is characterized by its sigmoid shape and ensures that the output is bounded within the specified range.
The interconnected system must satisfy a persistent objective, ensuring it reaches and maintains a state within the consensus region defined by:
\begin{equation}
\varphi = \bigvee_{\phi \in [0;31]} \Biggl(\bigwedge_{i=1}^N (\|x_i - \phi\| \leq 2.5)\Biggr).
\end{equation}
This signifies that all agents have states $x_i$ within a neighborhood of a common value $\phi$. Once the consensus region is attained, the value of $\phi$ may still change over time as the interconnected
system evolves. If $\phi$ was a known and fixed constant, achieving this objective would be trivially decomposable into 
$N$ distinct tasks, and a controller could be synthesized for each subsystem independently. However, because $\phi$ may vary and is not predetermined, it becomes non-trivial to synthesize a controller for each subsystem 
$\Xi_i$ individually, particularly if the controllers only have access to the local state 
$x_i$. The reach and remain objective can be succinctly expressed in a temporal logic formula as $\Diamond\square\varphi$ \cite{baier2008principles}.

Two inherent subsystem's properties exhibit challenges in achieving $\Diamond\square\varphi$. Firstly, with
fixed values of $u_i = 0$ and $w = 0$, system \eqref{ex2_f} has an unstable equilibrium at $x_i^*=16$ and two stable equilibria at $x_i^*=1.45$ and $x_i^*=30.5$. This bimodal behavior introduces difficulties, particularly, when the subsystems have initial states both above and below $16$, leading to divergence. Secondly, the interaction between the subsystems adds complexity. The concrete interconnection map is defined as the average state among all subsystems: 
\begin{equation}
\label{ex2_w}
    w=\frac{1}{N}\sum_{i=1}^N x_i,
\end{equation}
where $w\in[0,32]$, $\forall [x_1;\ldots;x_N]\in\prod_{i=1}^N \mathcal{X}_i$.  If $u_i(k) = 0$,
then the term $0.1(x_{i}-w)$ acts to push the
state $x_i$ away from the average, destabilizing the consensus region $\varphi$.

Here, we consider a network of $8$ subsystems described by \eqref{ex2_f} and interconnected with \eqref{ex2_w}. Hence, the exogenous input $w$ relies on $8$ values (cf. Fig. \ref{figdec} (top)), making the construction of the interconnection abstraction in a monolithic way very difficult. To address this, we collect $100$ and $20$ data points (denoted as $\mathcal{N}_{C_i}$ for the subsystems and $\Bar{\mathcal{N}}_I$ for the interconnection) using the models in \eqref{ex2_f} and \eqref{ex2_w}. The abstract sets $\hat{\mathcal{X}}_i$ and $\hat{\mathcal{U}}$ were generated using grid parameters $\eta_{x_i}=\eta_w=0.5$, while the abstract external input set for subsystems was defined as $\hat U_i=\left[U_i\right]_1$. Furthermore, we estimated the Lipschitz constants as $\mathcal{L}_{x_i}(\hat u)=\mathcal{L}_{w_i}(\hat u)=1.599$ for all $\hat u\in \hat U$, and $\mathcal{L}_{\mathcal{M}}=\frac{1}{8}$ (cf. Remark \ref{rmk_assump}). To construct the interconnection abstraction $\hat{\mathcal{I}}=\big(\prod^8_{i=1}\hat{\mathcal{X}}_i,\hat{\mathcal{U}},\hat{\mathcal{M}}\big)$, we solved \eqref{lasso} using $\Bar{\mathcal{N}}_I$ sampled  input-output data points obtained from \eqref{ex2_w}. Upon obtaining $\widebar{\mathcal{M}}\in\mathbb{R}^8$ picking $\sigma=4$, we applied a decomposition of the form $M_1:=4\widebar{\mathcal{M}}\cdot[\mathbf{1}_{8}~\;\mathbf{1}_{8}]$ and $M_2:=2\widebar{\mathcal{M}}\cdot[\mathbf{1}_{8}~\;\mathbf{1}_{8}~\;\mathbf{1}_{8}~\;\mathbf{1}_{8}]$ (cf. Algorithm \ref{alg_dec}) to introduce the following intermediate variables. 
For any $[x_1;\ldots;x_8]\in\prod^8_{i=1}\hat{\mathcal{X}}_i$, we define:
\begin{equation}
    \begin{split}
        z_1&=M_2\cdot[x_1;x_2;x_3;x_4]^\top\in[\mathcal{X}_i]_2,\\
    z_2&=M_2\cdot[x_5;x_6;x_7;x_8]^\top\in[\mathcal{X}_i]_2,\\
        w&=M_1\cdot[z_{1};z_{2}]^\top\in\hat{\mathcal{U}}.
    \end{split}
\end{equation}

\begin{figure}[!ht]
\vspace{-0.5cm}
    \centering
    \includegraphics[width=12.0cm]{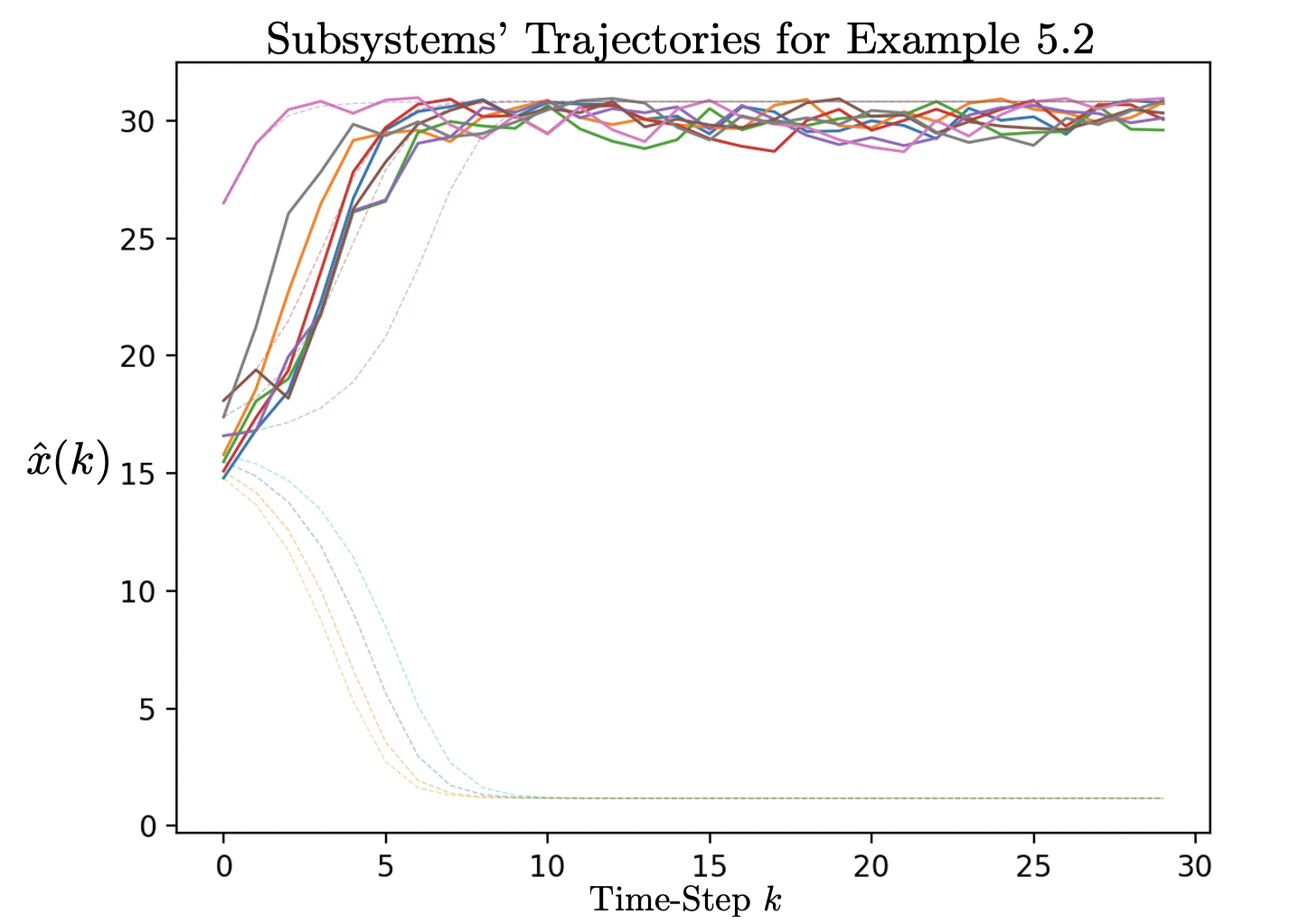}
    \includegraphics[width=12.0cm]{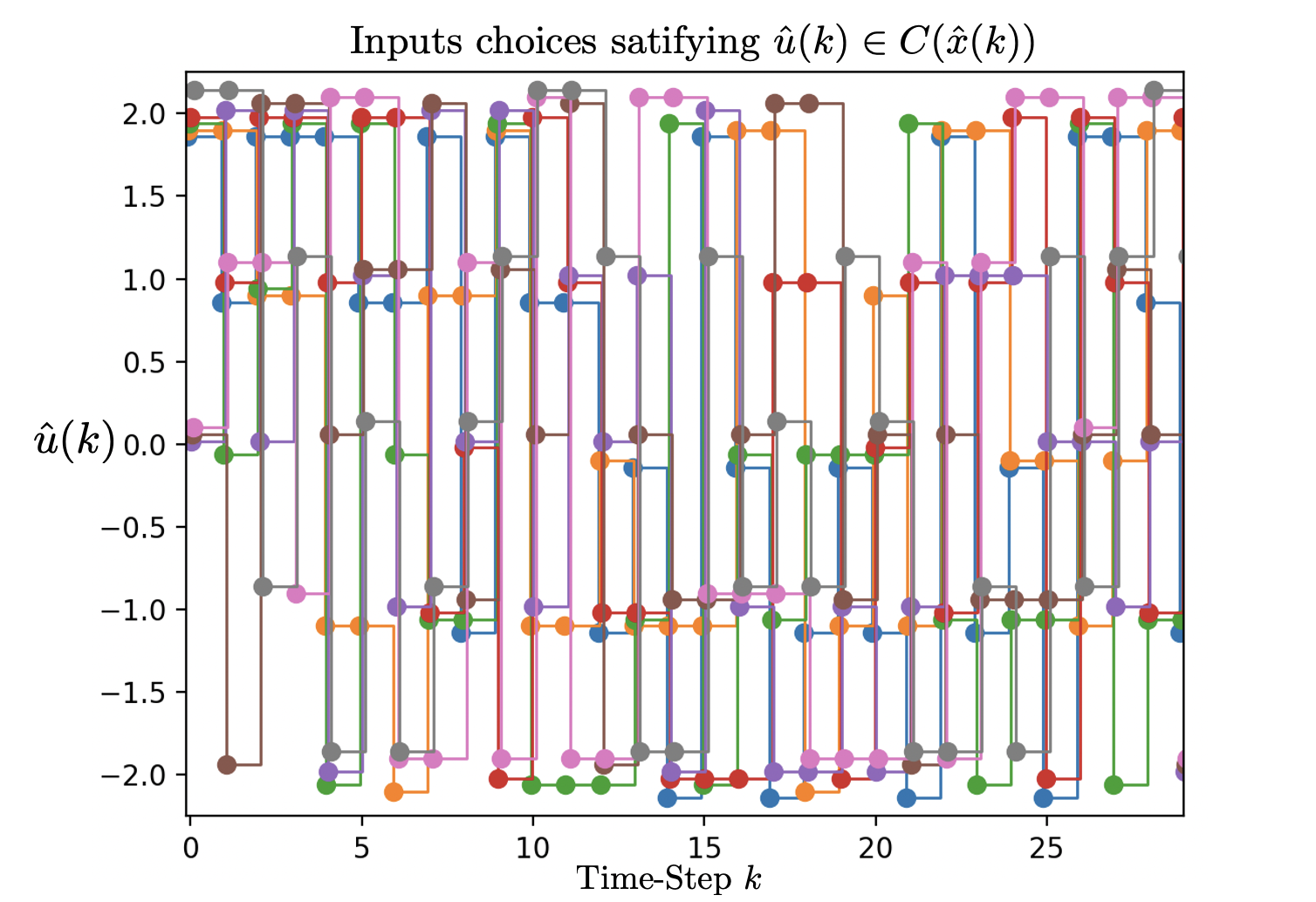}
    \caption{The top sub-figure depicts two 8-dimensional trajectories, which are illustrated as two sets of eight scalar-valued trajectories. Each line represents a trajectory $x_i(k)$ for $i\in[1;8]$. Solid lines correspond to a controller enforcing $\Diamond\square\varphi$, while dashed lines represent an inactive controller with $u_i(k)=0$ for all systems $i$ and time steps $k$. Both sets of trajectories initiate from the same initial state $x[0]=[14.8;15.8;15.5;15.1;16.6;18.1;26.5;17.4]$, which lies outside the consensus region $\varphi$ and supports the bimodal property of \eqref{ex2_f}. The bottom sub-figure depicts the piecewise constant control inputs required to enforce $\Diamond\square\varphi$. The depiction is with slight vertical and horizontal perturbations for visual clarity.}
    \label{fig4}
\end{figure}
The abstract interconnected system has $1.406\times 10^{12}$ states and the consensus region $\varphi$ encompasses $9.49\times 10^6$ discrete states. The processes of abstracting and stacking of the abstraction BDD representations of the subsystems took $300.85s$, and constructing the interconnection abstraction took $3.89s$. Then, we synthesized a controller on the interconnected abstraction, with the objective of persistently maintaining the system within the 
$\varphi$ region. At each state $\hat x\in\prod_{i=1}^8 \hat{\mathcal{X}}_i$, the controller provides a set of permissible  inputs $C(\hat x) \subseteq \hat U$. Ensuring 
$\hat u(k)\in C(\hat x(k))$ for all time steps guarantees satisfaction of the specification for the interconnected system. As demonstrated, the controller effectively ensures the persistence of the interconnected system's trajectory within region 
$\varphi$, as depicted in Fig. \ref{fig4}. The controller's domain encompasses $70.6\times 10^9$ states. Fig. \ref{fig4} illustrates two sets of trajectories corresponding to an inactive controller and an enforcing controller, which respectively violate and ensure satisfaction of 
$\Diamond\square\varphi$. In Fig. \ref{fig4}, control inputs $\hat u(k)$ are
randomly selected from $C(\hat x(k))$.

\subsection{Stabilizing a Network of Tanks in Target Regions}
\label{sec53}
\begin{figure}[!ht]
\vspace{-0.4cm}
    \centering
    \includegraphics[width=12.0cm]{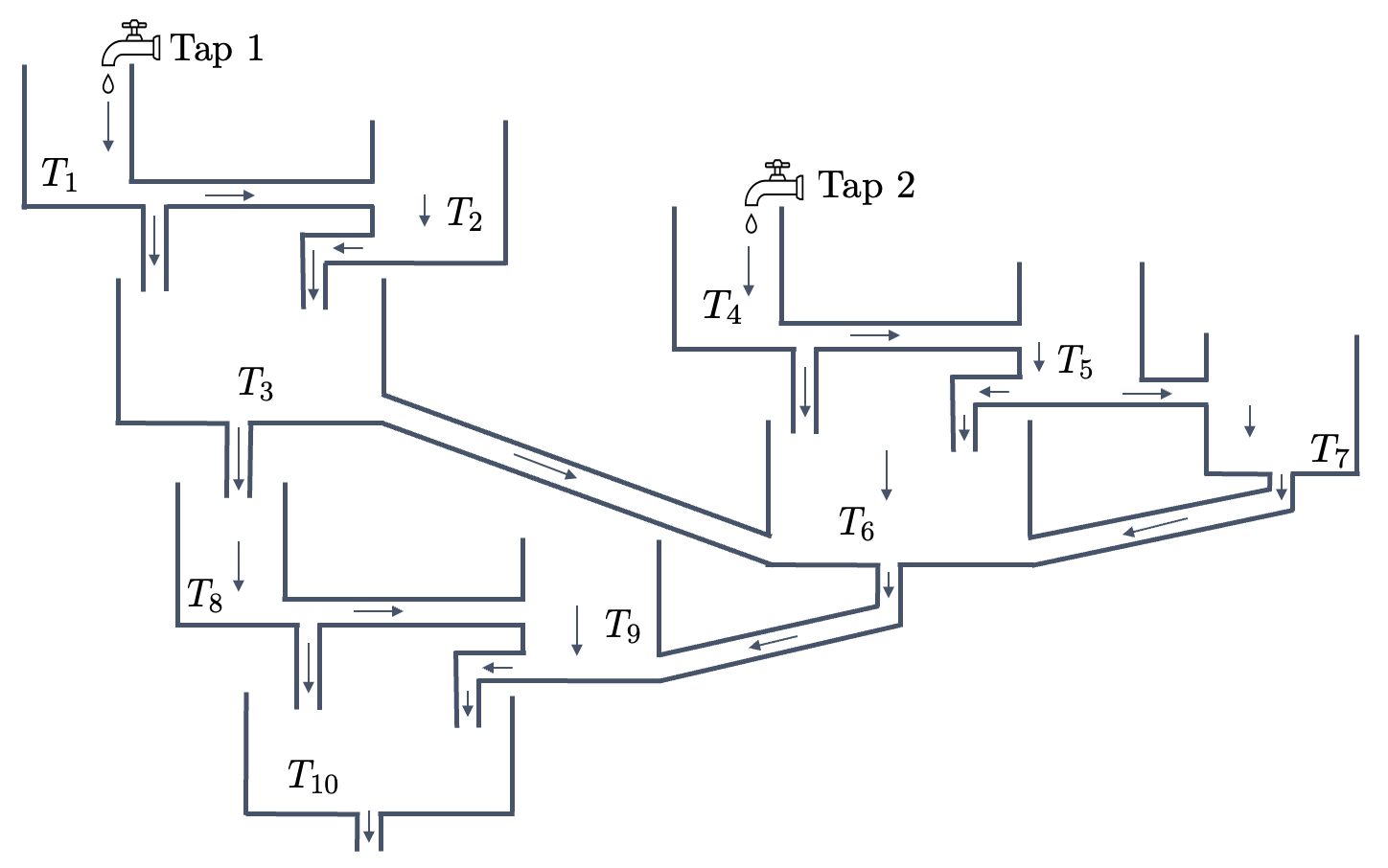}
    \caption{The configuration of $10$-tank models, represented by dt-CSs in \eqref{tank_dtcs}.} 
    \label{tank_confg}
    \vspace{0.2cm}
\end{figure}
Our final example involves a network of $10$ tanks, where each tank $T_1,\ldots,T_{10}$ forms a subsystem arranged in a cascade configuration, as shown in Fig. \ref{tank_confg}. Each tank is characterized by state sets $\mathcal{X}_i=[1,100]$, input sets $U_i=[0,10]$, and internal input sets $\mathcal{U}_i=[1,100]$, where $i\in[1;10]$. The dynamics governing each subsystem $\Xi_i$ are defined as:
\begin{align}
    \label{tank_dtcs}
    x_i(k+1)&=\Bigg[-\frac{\tau_s}{2}+\sqrt{\frac{\tau_s^2}{4}+x_i(k)+\tau_s\Big(u_i(k)+w_i(k+1)\Big)}~\;\Bigg]^2\notag\\
    x_j(k+1)&=\Bigg[-\frac{\tau_s}{2}+\sqrt{\frac{\tau_s^2}{4}+x_j(k)+\tau_s w_j(k+1)}~\;\Bigg]^2,
\end{align}
where $(i,j)\in\{(1,2),(1,3),(2,3),(3,6),(3,8),(4,5),(4,6),$ $(5,6),(5,7),(6,9),(7,6),(8,9),(8,10),(9,10)\}$. The sampling time is $\tau_s=0.5$ seconds. For any $i\in[1; 10]$, the state $x_i(k)\in\mathcal{X}_i$ and the internal input $w_i(k)\in\mathcal{U}_i$ represent the fluid level and the outflow
rate of the $i$-th tank at time $k\in\mathbb{N}$, respectively. The inflow rate $u_1\in U_1$ from tap $1$ and $u_4\in U_4$ from tap $2$ supply the first tank $T_1$ and the fourth tank $T_4$, respectively, as the first and fourth components of the networked system's input, while other components takes value from set $\{0\}$. Similar to the previous example, the interconnected system must satisfy a persistent objective, ensuring it reaches and stays within a target region. This reach-and-stay objective is expressed concisely using the temporal logic formula $\Diamond\square\psi$ \cite{baier2008principles} where \begin{equation}
\psi = \bigwedge_{i=1}^{10} \Big(\|x_i - 30\| \leq 20\Big).
\end{equation}
Our goal is to design a controller that ensures the interconnected system eventually reaches the target region $[10,50]^{10}$ and remains there afterwards.

The concrete interconnection map is defined as follows:
\begin{equation}
    \label{tank_inter}
    \begin{split}
    w(k)=\Bigg[
        0~\;;\sqrt{x_1(k)}~\;;\frac{1}{2}\Big(\sqrt{x_1(k)}+\sqrt{x_2(k)}\Big)~\;;0~\;;\sqrt{x_4(k)};\\~\;\;\frac{1}{4}\Big(\sqrt{x_3(k)}+\sqrt{x_4(k)}+\sqrt{x_5(k)}+\sqrt{x_7(k)}\Big)~\;;\sqrt{x_5(k)};\\
        ~\;\;\sqrt{x_3(k)}~\;;\frac{1}{2}\Big(\sqrt{x_6(k)}+\sqrt{x_8(k)}\Big)~\;;\frac{1}{2}\Big(\sqrt{x_8(k)}+\sqrt{x_9(k)}\Big)\Bigg],
        \end{split}
\end{equation}
where $w(k)=[w_1(k);w_2(k);\ldots;w_{10}(k)]\in\mathcal{U}^\mathcal{I}$, $\forall [x_1;\ldots;x_{10}]\in\mathcal{X}^\mathcal{I}$. 
\begin{figure}[ht]
\centering   
\subfigure{\label{tank_declf}\includegraphics[width=150mm]{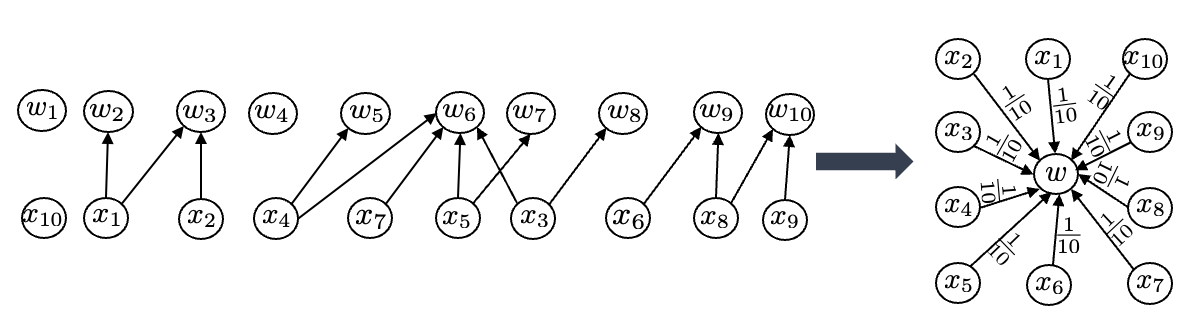}}
\subfigure{\label{tank_decrght}\includegraphics[width=150mm]{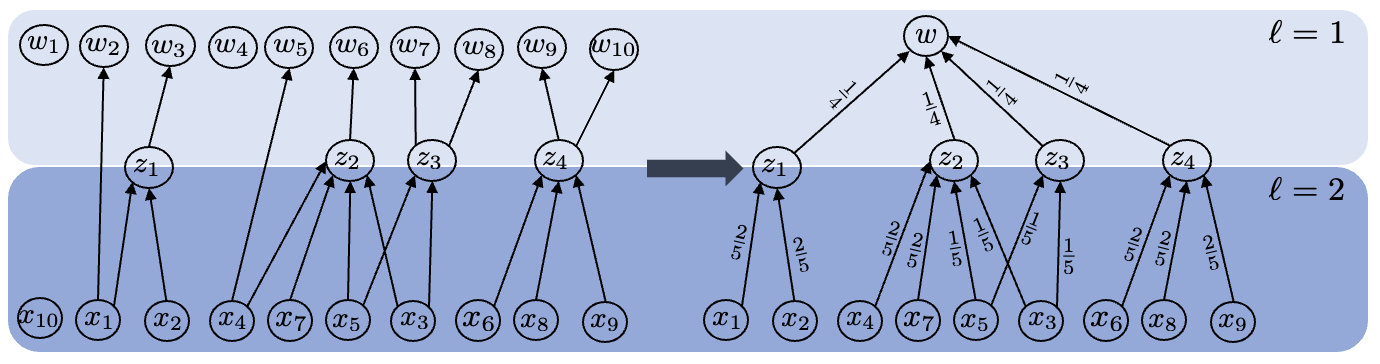}}
\caption{A WDAG for \eqref{tank_inter} with the application of Algorithm \ref{alg_dec} where $\sigma=4$.}
\label{tank_figdec}
\end{figure}
Fig. \ref{tank_figdec} (top) illustrates the connection between the exogenous inputs and the subsystem states. We observe that $x_{10}$ lacks a connecting edge in Fig. \ref{tank_figdec}, as the outflow from tank $T_{10}$ does not feed into any other tank in this experiment. This is further confirmed by the concrete interconnection map $\mathcal{M}$ in \eqref{tank_inter}. We collected 150 data points for the subsystems (denoted by \( \mathcal{N}_{C_i} \)) and 80 data points for the interconnection (\( \Bar{\mathcal{N}}_I \)) using the models in \eqref{tank_dtcs} and \eqref{tank_inter}. The abstract sets $\hat{\mathcal{X}}_i$ and $\hat{\mathcal{U}}$ were generated using grid parameters $\eta_{x_i}=\eta_{w_i}=2.5$, while the abstract external input set for the subsystems was defined as $\hat U_i=\left[U_i\right]_{1.5}$. Furthermore, we estimated the Lipschitz constants as $\mathcal{L}_{x_i}(\hat u)=0.9083$, $\mathcal{L}_{w_i}(\hat u)=0.9988$ for all $\hat u\in \hat U$, and $\mathcal{L}_{\mathcal{M}}=0.5$ (cf. Remark \ref{rmk_assump}). To construct the interconnection abstraction $\hat{\mathcal{I}}=\big(\prod^{10}_{i=1}\hat{\mathcal{X}}_i,\hat{\mathcal{U}},\hat{\mathcal{M}}\big)$, we solved \eqref{lasso} using $\Bar{\mathcal{N}}_I$ sampled input-output data points obtained from \eqref{tank_inter}. Upon obtaining $\widebar{\mathcal{M}}\in\mathbb{R}^{10\times 10}$ of the form 
\begin{equation}
    \label{tank_M_bar}
    \begin{bmatrix}
        M_{1}&\mathbf{0}_{4\times6}\\
        \mathbf{0}_{7\times2}&\wp
    \end{bmatrix},
\end{equation}
by picking $\sigma=4$, where $\wp\in\mathbb{R}^{6\times 8}$, we applied a decomposition of the form 
\begin{equation*}
    \begin{split}
      M_{2_1}&:=[\mathbf{I}_{2};\mathbf{0}_{2\times4}]\cdot\wp\cdot[\mathbf{I}_{3};\mathbf{0}_{3\times5}]^\top, \\
      M_{2_2}&:=[\mathbf{I}_{2};\mathbf{0}_{2\times4}]\cdot\wp\cdot[\mathbf{0}_{2\times4};\mathbf{I}_{2};\mathbf{0}_{2\times2}]^\top,\\
      M_3&:=[\mathbf{0}_{2\times2};\mathbf{I}_{2};\mathbf{0}_{2\times2}]\cdot\wp\cdot[\mathbf{I}_{3};\mathbf{0}_{3\times5}]^\top,\\
      M_{4_1}&:=[\mathbf{0}_{2\times4};\mathbf{I}_{2}]\cdot\wp\cdot[\mathbf{0}_{2\times4};\mathbf{I}_{2};\mathbf{0}_{2\times2}]^\top\text{ and}\\
      M_{4_2}&:=[\mathbf{0}_{2\times4};\mathbf{I}_{2}]\cdot\wp\cdot[\mathbf{0}_{3\times5};\mathbf{I}_{3}]^\top,\\
    \end{split}
\end{equation*}
(cf. Algorithm \ref{alg_dec}) to introduce the following intermediate variables. For any $[x_1;\ldots;x_{10}]\in\prod^{10}_{i=1}\hat{\mathcal{X}}_i$, we define:
\begin{equation}
    \begin{split}
z_1&=M_1\cdot[x_1;x_2]^\top\in\Big([\mathcal{X}_i]_5\Big)^3,\\
    z_2&=M_{2_1}\cdot[x_3;x_4;x_5]^\top+M_{2_2}\cdot[0;x_7]^\top\in\Big([\mathcal{X}_i]_5\Big)^2,\\
    z_3&=M_3\cdot[x_5;0;x_3]^\top\in\Big([\mathcal{X}_i]_5\Big)^2,\\
    z_4&=M_{4_1}\cdot[x_6;0]^\top+M_{4_2}\cdot[x_8;x_9;0]^\top\in\Big([\mathcal{X}_i]_5\Big)^2,\\
        w&=[z_{1};0;z_{2};z_3;z_4]^\top\in\hat{\mathcal{U}}.
    \end{split}
\end{equation}

\begin{figure}[!ht]
\vspace{-0.5cm}
    \centering
    \includegraphics[width=12.0cm]{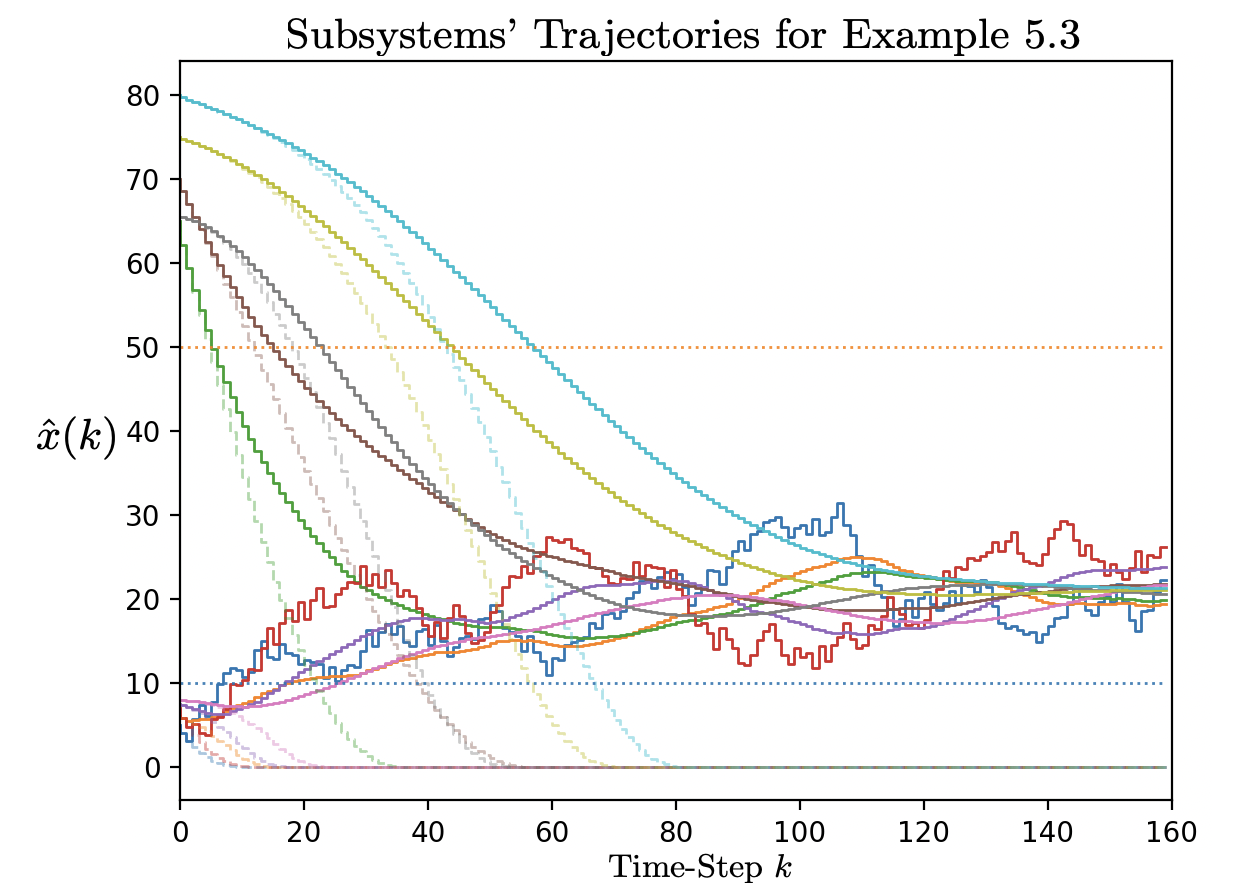}
    \includegraphics[width=12.0cm]{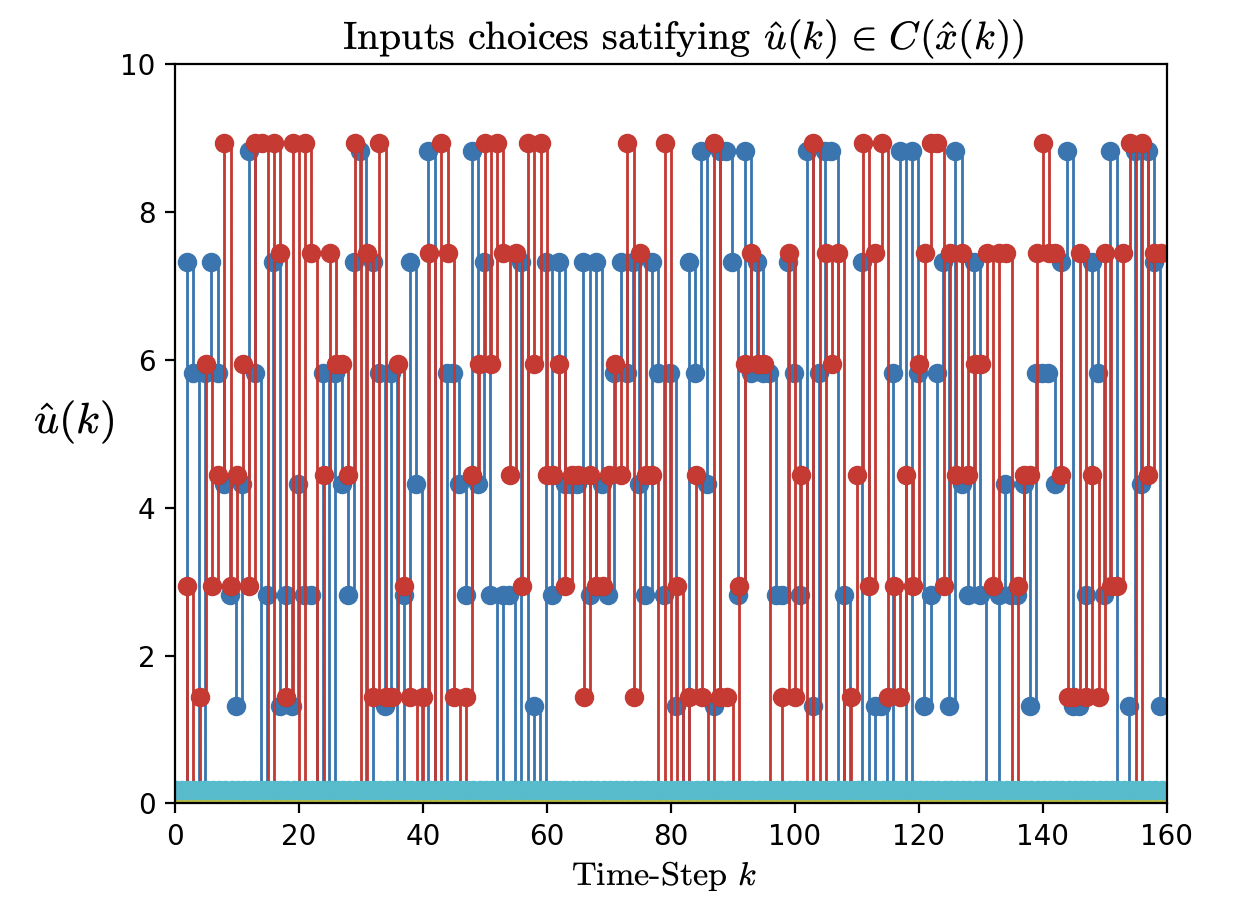}
    \caption{The top sub-figure shows two $10$-dimensional trajectories, displayed as two sets of ten scalar-valued trajectories. Each line represents a trajectory $x_i(k)$ for $i\in[1;10]$. Solid lines correspond to a controller enforcing $\Diamond\square\psi$, while dashed lines represent an inactive controller with $u_i(k)=0$ for all systems $i$ and time steps $k$. Both sets of trajectories start from the same initial state $x[0]=[5;6;65;7;7.5;70;8;65.5;75;80]$, which is outside the target region $\psi$. The bottom sub-figure shows the piecewise constant control inputs required to enforce $\Diamond\square\psi$, with slight vertical and horizontal perturbations added for visual clarity.}
    \label{fig5}
\end{figure}
The abstract interconnected system contains \( 1.34 \times 10^{16} \) states, with the target region \( \psi \) covering \( 2.016 \times 10^{12} \) discrete states. Abstracting and stacking the BDD representations of the subsystems took 262.51 seconds, while constructing the interconnection abstraction required 58.76 seconds. A controller was then synthesized for the interconnected abstraction to reach and maintain the system within the \( \psi \) region. As shown in Fig. \ref{fig5}, the controller effectively enforces the desired specification. The controller's domain comprises \( 6.729 \times 10^{14} \) states. Fig. \ref{fig5} illustrates two sets of trajectories: one without the controller, which violates \( \Diamond\square\psi \), and one with the designed controller, which ensures satisfaction. Without the controller (\emph{i.e.,} $u_i(k)=0~\forall i\in[1;10]$), all tanks eventually empty, whereas the designed controller stabilizes the fluid levels within the target region \( \psi \).

As noted in Remark \ref{rem_conserve}, we set \( \vartheta_1(\hat{x}, \hat{u}, \hat{w}) = \mathcal{L}_{x_i}(\hat{u}) \) and \( \vartheta_2(\hat{x}, \hat{u}, \hat{w}) = \mathcal{L}_{w_i}(\hat{u}) \eta_{w_i} \) and use the resulting growth bound to construct the subsystem abstraction. The resulting abstract interconnected system is highly conservative, leading to synthesized controllers with significantly smaller winning domains. In this case study, the synthesized controller's domain for the conservative abstraction is reduced by \( 99.58\% \) compared to the domain obtained from the less conservative abstraction constructed using the growth bound from solving SCP \eqref{c_scp} for the subsystems.

Additionally, due to the high dimensionality of the interconnected system in this case study, attempting to use data-driven monolithic approaches from the literature, such as \cite{kazemi2022data}, \cite{gruber2017sparsity} and \cite{ajeleye2023data}, to construct a finite abstraction resulted in a computation timeout.

\appendix

\renewcommand{\thetheorem}{\Alph{section}.\arabic{theorem}}
\renewcommand{\thelemma}{\Alph{section}.\arabic{lemma}}
\renewcommand{\theremark}{\Alph{section}.\arabic{remark}}

\section{Data-Driven Estimation of Lipschitz Constants}\label{apdx1}
In this section, we leverage the results presented in \cite{wood1996estimation} to introduce the following algorithm for estimating the Lipschitz constants of the subsystems' dynamics and the interconnection map. This is accomplished using a finite set of data points collected from the black-box representation of the system.
\begin{algorithm}[ht!]
	\caption{Estimation of Lipschitz Constants}
 \label{alg_lip}
 \begin{algorithmic}[1]
\REQUIRE dt-CS $\Xi$, $\hat{U}$ and $\mathcal{X}$
    \STATE Select $J,L\in\mathbb{N}_{\ge1}$ with points $x_l,\bar x_l\in \mathcal{X}$ such that $\Vert x_l-\bar x_l'\Vert\le\delta$, $\forall l\in[1;L]$ where $\delta>0$
    \FOR {$\hat u\in \hat U$}
    \FOR {$j\in[1;J]$}
    \FOR {$l\in[1;L]$}
    \STATE Simulate the dt-CS $\Xi$ from initial states $x_l$ and $\bar x_l$ in one time-step under input $\hat u$ to obtain $x_l'$ and $\bar x_l'$, respectively
    \STATE Compute slope $\Delta_l:=\dfrac{\Vert x_l'-\bar x_l'\Vert}{\Vert x_l-\bar x_l\Vert}$
    \ENDFOR
    \STATE Obtain the maximum slope $\Delta^*_j:=\max\{\Delta_1,\ldots,\Delta_L\}$
    \ENDFOR
    \STATE Fit $\{\Delta^*_1,\ldots,\Delta^*_J\}$ into a Reverse Weibull distribution \cite{wood1996estimation} to obtain the parameters termed location, scale and shape
    \ENDFOR
    \STATE The estimated Lipschitz constant $\mathcal{L}_x(\hat u)$ is the obtained \emph{location} parameter
\ENSURE $\mathcal{L}_x(\hat u)$ $\forall \hat u\in \hat U$
\end{algorithmic}
\end{algorithm}

Through the implementation of Algorithm \ref{alg_lip}, the Lemma \ref{lem_apd}, sourced from \cite{wood1996estimation}, guarantees that the estimated Lipschitz constant converges towards its actual value in the limit. 
\begin{lemma}
\label{lem_apd}
    Consider a dt-CS $\Xi$ with partially unknown transition function. By applying Algorithm \ref{alg_lip}, the estimated Lipschitz constant $\mathcal{L}_x(\hat u)$ 
    converges to the actual value if and only if $\delta$ gets arbitrarily small and $J,L$ becomes very large.
\end{lemma}
\begin{remark}
    Note that we do not consider any confidence bounds when estimating the Lipschitz constants in our approach. Instead, we choose a very small value for \( \delta \) and large values for \( J \) and \( L \) to ensure that Algorithm \ref{alg_lip} provides a reliable estimate of the Lipschitz constant. Additionally, Algorithm \ref{alg_lip} can be adjusted to estimate the Lipschitz constant \( \mathcal{L}_w \) by setting \( \mathcal{X} \) to \( \mathcal{U} \). Similarly, to estimate the Lipschitz constant \( \mathcal{L}_{\mathcal{M}} \) for an unknown nonlinear interconnection map, we can modify Algorithm \ref{alg_lip} by setting \( \hat{U} = \emptyset \) and \( \mathcal{X} \) to \( \mathcal{X}^\mathcal{I} \). In this case, Line 5 becomes: simulate the black-box representation of \( \mathcal{M} \) with inputs \( x_l \) and \( \bar{x}_l \) to obtain outputs \( x_l' \) and \( \bar{x}_l' \), respectively.
\end{remark}

\section{Proof of Theorem \ref{sub_abs_thrm}}
\label{apdx3}
\begin{proof}
To establish the proof of Theorem \ref{sub_abs_thrm}, we need to show that the two conditions in Definition \ref{asf} are satisfied for the dt-CS $\widehat{\Xi}$ and $\Xi$. We begin by verifying the first condition. Consider $x\in\Phi_{\eta_x/2}(\hat x)$ and $w\in\Phi_{\eta_w/2}(\hat w)$, where $\hat x\in \hat{\mathcal{X}}$ and $\hat w\in \hat{\mathcal{U}}$. From Definition \ref{sub_abs}, let $\mathscr{X}:=(f(\hat x,\hat u,\hat w)\oplus[-q',q'])\cap\Phi_{\eta_x/2}(\hat x')$, where $q'=\kappa(\eta_x,\hat x,\hat u,\hat w)$. Since $\widehat{\Xi}$ is an abstraction of $\Xi$, it follows from Definition \ref{sub_abs} that if $\hat x\in \hat{\mathcal{X}}\setminus \mathscr{X}$, then $U_{\Xi_2}(\hat x,\hat w)=\emptyset$. Conversely, if $x\in\mathscr{X}$, then $U_{\Xi_2}(\hat x,\hat w)\subseteq U_{\Xi_1}(x,w)$, satisfying the first condition of Definition \ref{asf}. Next, we verify the second condition. Suppose that $\hat x,\hat x'\in\hat{\mathcal{X}}$ and $u\in U_{\Xi_2}(\hat x,\hat w)$. Given that $\hat{\Xi}$ is an abstraction of $\Xi$, assuming $\hat x\in\mathscr{X}$ implies $\hat x\in (c\oplus[-r,r])$ for some $c\in\mathbb{R}^{\mathrm{dim}(\mathcal{X})}$, $r\in \mathbb{R}^{\mathrm{dim}(\mathcal{X})}_{>0}$. If $x\in\Phi_{\eta_x/2}(\hat x)$ and $f(x,u,w)\in\Phi_{\eta_x/2}(\hat x')$, then, by Assumption \ref{assume2}, we have $x\in(c\oplus[-r,r])$. Consequently, $\hat x'\in\big(f(c,u,w)\oplus[-r',r']\big)$ for some $r'\in\mathbb{R}^{\mathrm{dim}(\mathcal{X})}_{>0}$. Thus, since $\widehat\Xi$ is an abstraction, $\hat x'\in\hat f(\hat x,u,\hat w)$, verifying the second condition of Definition \ref{asf}. This completes the proof.
\end{proof}

\section{Proof of Lemma \ref{lema1_c}}
\label{apdx2}
\begin{proof}
Using \eqref{eqlp2}, it holds that $\forall \hat x\in \hat{\mathcal{X}}$ with $x_1,x_2\in\Phi_{\eta_x/2}(\hat x)$ and given $(\hat u,\hat w)\in \hat U\times\hat{\mathcal{U}}$ with $w_1,w_2\in\Phi_{\eta_w/2}(\hat w)$,
    \begin{equation}
    \label{apd1}
    \Vert f(x_1,\hat u,w_1)-f(x_2,\hat u,w_2)\Vert\le \mathcal{L}_x(\hat u)\Vert x_1-x_2\Vert + \mathcal{L}_w(\hat u)\Vert w_1-w_2\Vert.
    \end{equation}
    Consider $\kappa_\vartheta$ as in \eqref{k_c_para}, then using $\vartheta_1(\hat x,\hat u, \hat w)=\mathcal{L}_x(\hat u) \mathbf{1}_{\mathrm{dim}(\mathcal{X}) \times \mathrm{dim}(\mathcal{X})}$ and $\vartheta_2(\hat x,\hat u, \hat w)=\mathcal{L}_w(\hat u)\eta_w$ in \eqref{k_c_para} and then applying the infinity norm on the resulting inequality, one obtains
    \begin{equation}
    \label{apd2}
        \begin{split}
        \Vert\kappa_\vartheta(|x_1-x_2|,\hat x,\hat u,\hat w)\Vert&\le\Vert\vartheta_1(\hat x,\hat u, \hat w)\Vert\,\Vert x_1-x_2\Vert + \Vert\vartheta_2(\hat x,\hat u, \hat w)\Vert\\
        &\le\mathcal{L}_x(\hat u)\Vert x_1-x_2\Vert + \mathcal{L}_w(\hat u)\Vert\eta_w\Vert.\\
        \end{split}
    \end{equation}
    By subtracting \eqref{apd2} from \eqref{apd1}, one has
    $$\Vert f(x_1,\hat u,w_1)-f(x_2,\hat u,w_2)\Vert\le\Vert\kappa_\vartheta(|x_1-x_2|,\hat x,\hat u,\hat w)\Vert,$$
    which concludes the proof.
\end{proof}

\bibliographystyle{alpha}
\bibliography{biblio.bib}

\newcommand{\etalchar}[1]{$^{#1}$}
\begin{thebibliography}{FQMV17}

\bibitem[ALZ23]{ajeleye2023data}
Daniel Ajeleye, Abolfazl Lavaei, and Majid Zamani.
\newblock Data-driven controller synthesis via finite abstractions with formal guarantees.
\newblock {\em IEEE Control Systems Letters}, 7:3453--3458, 2023.

\bibitem[AMP22]{ajeleye2022output}
Daniel~Ajedamola Ajeleye, Tommaso Masciulli, and Giordano Pola.
\newblock Output feedback control of nondeterministic finite--state systems with reach--avoid specifications.
\newblock In {\em 2022 30th Mediterranean Conference on Control and Automation (MED)}, pages 1012--1017. IEEE, 2022.

\bibitem[AZ24]{ajeleye2024data}
Daniel Ajeleye and Majid Zamani.
\newblock Data-driven controller synthesis via co-b{\"u}chi barrier certificates with formal guarantees.
\newblock {\em IEEE Control Systems Letters}, 2024.

\bibitem[BK08]{baier2008principles}
Christel Baier and Joost-Pieter Katoen.
\newblock {\em Principles of model checking}.
\newblock MIT press, 2008.

\bibitem[Bry86]{bryant1986graph}
Randal~E Bryant.
\newblock Graph-based algorithms for boolean function manipulation.
\newblock {\em Computers, IEEE Transactions on}, 100(8):677--691, 1986.

\bibitem[CC06]{calafiore2006scenario}
Giuseppe~Carlo Calafiore and Marco~C Campi.
\newblock The scenario approach to robust control design.
\newblock {\em IEEE Transactions on automatic control}, 51(5):742--753, 2006.

\bibitem[CPMJ22]{coppola2022data}
Rudi Coppola, Andrea Peruffo, and Manuel Mazo~Jr.
\newblock Data-driven abstractions for verification of deterministic systems.
\newblock {\em arXiv preprint arXiv:2211.01793}, 2022.

\bibitem[FQMV17]{fan2017dryvr}
Chuchu Fan, Bolun Qi, Sayan Mitra, and Mahesh Viswanathan.
\newblock Dry{VR}: Data-driven verification and compositional reasoning for automotive systems.
\newblock In {\em International Conference on Computer Aided Verification}, pages 441--461. Springer, 2017.

\bibitem[FS18]{fattahi2018data}
Salar Fattahi and Somayeh Sojoudi.
\newblock Data-driven sparse system identification.
\newblock In {\em 2018 56th Annual Allerton Conference on Communication, Control, and Computing (Allerton)}, pages 462--469. IEEE, 2018.

\bibitem[GKA17]{gruber2017sparsity}
Felix Gruber, Eric~S Kim, and Murat Arcak.
\newblock Sparsity-aware finite abstraction.
\newblock In {\em 2017 IEEE 56th Annual Conference on Decision and Control (CDC)}, pages 2366--2371. IEEE, 2017.

\bibitem[HAT17]{hussien2017abstracting}
Omar Hussien, Aaron Ames, and Paulo Tabuada.
\newblock Abstracting partially feedback linearizable systems compositionally.
\newblock {\em IEEE Control Systems Letters}, 1(2):227--232, 2017.

\bibitem[HRC23]{huang2023sample}
Julien~Walden Huang, Stephen~J Roberts, and Jan-Peter Calliess.
\newblock On the sample complexity of lipschitz constant estimation.
\newblock {\em Transactions on Machine Learning Research}, 2023.

\bibitem[HW13]{hou2013model}
Zhong-Sheng Hou and Zhuo Wang.
\newblock From model-based control to data-driven control: Survey, classification and perspective.
\newblock {\em Information Sciences}, 235:3--35, 2013.

\bibitem[KAZ18]{kim2018constructing}
Eric~S Kim, Murat Arcak, and Majid Zamani.
\newblock Constructing control system abstractions from modular components.
\newblock In {\em Proceedings of the 21st International Conference on Hybrid Systems: Computation and Control (part of CPS Week)}, pages 137--146, 2018.

\bibitem[KCOB21]{knuth2021planning}
Craig Knuth, Glen Chou, Necmiye Ozay, and Dmitry Berenson.
\newblock Planning with learned dynamics: Probabilistic guarantees on safety and reachability via lipschitz constants.
\newblock {\em IEEE Robotics and Automation Letters}, 6(3):5129--5136, 2021.

\bibitem[KMS{\etalchar{+}}22]{kazemi2022data}
Milad Kazemi, Rupak Majumdar, Mahmoud Salamati, Sadegh Soudjani, and Ben Wooding.
\newblock Data-driven abstraction-based control synthesis.
\newblock {\em arXiv preprint arXiv:2206.08069}, 2022.

\bibitem[Lav23]{lavaei2023symbolic}
Abolfazl Lavaei.
\newblock Symbolic abstractions with guarantees: A data-driven divide-and-conquer strategy.
\newblock In {\em 2023 62nd IEEE Conference on Decision and Control (CDC)}, pages 7994--7999. IEEE, 2023.

\bibitem[LF22]{lavaei2022data}
Abolfazl Lavaei and Emilio Frazzoli.
\newblock Data-driven synthesis of symbolic abstractions with guaranteed confidence.
\newblock {\em IEEE Control Systems Letters}, 7:253--258, 2022.

\bibitem[Lju98]{ljung1998system}
Lennart Ljung.
\newblock System identification.
\newblock In {\em Signal analysis and prediction}, pages 163--173. Springer, 1998.

\bibitem[MGF21]{makdesi2021efficient}
Anas Makdesi, Antoine Girard, and Laurent Fribourg.
\newblock Efficient data-driven abstraction of monotone systems with disturbances.
\newblock {\em IFAC-PapersOnLine}, 54(5):49--54, 2021.

\bibitem[MGG13]{mouelhi2013cosyma}
Sebti Mouelhi, Antoine Girard, and Gregor G{\"o}ssler.
\newblock Co{SyMA}: a tool for controller synthesis using multi-scale abstractions.
\newblock In {\em Proceedings of the 16th international conference on Hybrid systems: computation and control}, pages 83--88, 2013.

\bibitem[MGW17]{meyer2017compositional}
Pierre-Jean Meyer, Antoine Girard, and Emmanuel Witrant.
\newblock Compositional abstraction and safety synthesis using overlapping symbolic models.
\newblock {\em IEEE Transactions on Automatic Control}, 63(6):1835--1841, 2017.

\bibitem[MJDT10]{mazo2010pessoa}
Manuel Mazo~Jr, Anna Davitian, and Paulo Tabuada.
\newblock P{ESSOA}: A tool for embedded controller synthesis.
\newblock In {\em International conference on computer aided verification}, pages 566--569. Springer, 2010.

\bibitem[Nej23]{nejati2023formal}
Ameneh Nejati.
\newblock {\em Formal Verification and Control of Stochastic Hybrid Systems: Model-based and Data-driven Techniques}.
\newblock PhD thesis, Technische Universit{\"a}t M{\"u}nchen, 2023.

\bibitem[RWR16]{reissig2016feedback}
Gunther Reissig, Alexander Weber, and Matthias Rungger.
\newblock Feedback refinement relations for the synthesis of symbolic controllers.
\newblock {\em IEEE Transactions on Automatic Control}, 62(4):1781--1796, 2016.

\bibitem[RZ16]{rungger2016scots}
Matthias Rungger and Majid Zamani.
\newblock {SCOTS}: A tool for the synthesis of symbolic controllers.
\newblock In {\em Proceedings of the 19th international conference on hybrid systems: Computation and control}, pages 99--104, 2016.

\bibitem[SLSZ21]{salamati2021data}
Ali Salamati, Abolfazl Lavaei, Sadegh Soudjani, and Majid Zamani.
\newblock Data-driven safety verification of stochastic systems via barrier certificates.
\newblock In {\em 7th IFAC Conference on Analysis and Design of Hybrid Systems (ADHS 2021)}, volume~54, pages 7--12. Elsevier, 2021.

\bibitem[SLSZ24]{salamati2024data}
Ali Salamati, Abolfazl Lavaei, Sadegh Soudjani, and Majid Zamani.
\newblock Data-driven verification and synthesis of stochastic systems via barrier certificates.
\newblock {\em Automatica}, 159:111323, 2024.

\bibitem[Som97]{somenzi1997cudd}
Fabio Somenzi.
\newblock {CUDD}: {CU} decision diagram package.
\newblock {\em Public Software, University of Colorado}, 1997.

\bibitem[SZ22]{salamati2022data}
Ali Salamati and Majid Zamani.
\newblock Data-driven safety verification of stochastic systems via barrier certificates: A wait-and-judge approach.
\newblock In {\em Learning for Dynamics and Control Conference}, pages 441--452. PMLR, 2022.

\bibitem[Tab09]{tabuada2009verification}
Paulo Tabuada.
\newblock {\em Verification and control of hybrid systems: a symbolic approach}.
\newblock Springer Science \& Business Media, 2009.

\bibitem[Tib96]{tibshirani1996regression}
Robert Tibshirani.
\newblock Regression shrinkage and selection via the lasso.
\newblock {\em Journal of the Royal Statistical Society Series B: Statistical Methodology}, 58(1):267--288, 1996.

\bibitem[WZ96]{wood1996estimation}
GR~Wood and BP~Zhang.
\newblock Estimation of the lipschitz constant of a function.
\newblock {\em Journal of Global Optimization}, 8:91--103, 1996.

\bibitem[XZEL20]{xue2020pac}
Bai Xue, Miaomiao Zhang, Arvind Easwaran, and Qin Li.
\newblock {PAC} model checking of black-box continuous-time dynamical systems.
\newblock {\em IEEE Transactions on Computer-Aided Design of Integrated Circuits and Systems}, 39(11):3944--3955, 2020.

\end{thebibliography}
\end{document}